\documentclass[a4paper,11pt]{article}
\pdfoutput=1 

\usepackage{jcappub} 

\usepackage{amsmath}
\usepackage{mathtools}
\usepackage{xparse}

\newcommand{\ie}{i.\,e.~}

\newcommand{\eg}{e.\,g.~}

\newcommand{\wrt}{w.\,r.\,t.~}

\newcommand{\mi}{\mathrm{i}}
\newcommand{\mf}{\mathrm{f}}
\newcommand{\md}{\mathrm{d}}
\newcommand{\mm}{\mathrm{m}}
\newcommand{\mC}{\mathrm{C}}
\newcommand{\mS}{\mathrm{S}}
\newcommand{\mF}{\mathrm{F}}
\newcommand{\mg}{\mathrm{g}}

\newcommand{\mA}{\mathrm{A}}
\newcommand{\mB}{\mathrm{B}}
\newcommand{\mlin}{\mathrm{lin}}
\newcommand{\mpec}{\mathrm{pec}}
\newcommand{\mMC}{\mathrm{MC}}
\newcommand{\mSE}{\mathrm{SE}}
\newcommand{\e}{\mathrm{e}}

\newcommand{\gqq}{\mg_{\mathrm{qq}}}
\newcommand{\gqp}{\mg_{\mathrm{qp}}}
\newcommand{\gpq}{\mg_{\mathrm{pq}}}
\newcommand{\gpp}{\mg_{\mathrm{pp}}}

\newcommand{\ve}{\vec{e}}

\newcommand{\ini}[1]{#1^{(\mathrm{i})}}
\newcommand{\vini}[1]{\vec{#1}^{\,(\mathrm{i})}}
\newcommand{\tini}[1]{\boldsymbol{#1}^{(\mathrm{i})}}

\newcommand{\vx}{\vec{x}}

\newcommand{\tx}{\boldsymbol{x}}
\newcommand{\txi}{\tini{x}}

\newcommand{\qin}{\ini{q}}
\newcommand{\vq}{\vec{q}}
\newcommand{\vqi}{\vini{q}}
\newcommand{\tq}{\boldsymbol{q}}

\newcommand{\vp}{\vec{p}}
\newcommand{\vpi}{\vini{p}}
\newcommand{\tp}{\boldsymbol{p}}

\newcommand{\vupec}{\vec{u}_{\mpec}}
\newcommand{\vupeci}{\vini{u}_{\mpec}}

\newcommand{\vk}{\vec{k}}
\newcommand{\vh}{\vec{h}}

\newcommand{\vchi}{\vec{\chi}}
\newcommand{\vochi}{\hat{\vchi}}
\newcommand{\tchi}{\boldsymbol{\chi}}

\newcommand{\tYi}{\tini{Y}}

\newcommand{\mpd}{\bar{\rho}} 						
\newcommand{\dirac}{\delta_{\textsc{d}}}				
\newcommand{\heavi}{\Theta}						

\newcommand{\tJ}{\boldsymbol{J}}					
\newcommand{\tK}{\boldsymbol{K}}					

\newcommand{\psd}[1]{f^{(#1)}}						

\NewDocumentCommand \Gf{o}{						
  \IfNoValueTF{#1}
    { G^{(0)} }   
  { G^{(0,#1)} }
  }

\newcommand{\iniPS}{\ini{P}_\delta}

\NewDocumentCommand \Gdd{o}{						
  \IfNoValueTF{#1}  
    { G_{\delta \delta} }   
  { G^{(#1)}_{\delta \delta}}
 }

\newcommand{\upi}[1]{\int \mathcal{D} #1 \,}				

\newcommand{\bpint}[3]{\int\limits_{#2}^{#3} \mathcal{D} #1 \,} 	

\newcommand{\usi}[1]{\int \mathrm{d} #1 \,}				

\newcommand{\bsi}[3]{\int\limits_{#2}^{#3} \mathrm{d} #1 \,}		

\newcommand{\umi}[2]{\int \mathrm{d}^{#1} #2 \,}			

\newcommand{\fmi}[2]{\int \frac{\mathrm{d}^{#1} #2}{(2\pi)^{#1}} \,}    

\newcommand{\IC}{\int \mathrm{d} \Gamma_\mi \,}				

\newcommand{\tder}[2]{\frac{\mathrm{d} #1}{\mathrm{d} #2}}		

\newcommand{\fd}[2]{\frac{\delta #1}{\mi \delta #2}}			

\newcommand{\ave}[1]{\left\langle #1 \right\rangle}			

\newcommand{\nexp}[1]{\mathrm{exp}\left\{ #1 \right\}}			

\newcommand{\tens}[1]{\boldsymbol{#1}} 					

\newcommand{\cvector}[1]{\left(\begin{array}{c}#1\end{array}\right)}	

\renewcommand{\matrix}[2]{\left(\begin{array}{#1} #2\end{array}\right)}  

\usepackage{tikz}
\usepackage{pgfplots}
\pgfplotsset{compat=1.11}
\usepackage{xparse}

\usetikzlibrary{
positioning,
decorations.pathmorphing,
decorations.pathreplacing,
decorations.markings,
shapes.geometric,
calc
}

\newlength{\gs}
\newlength{\feyn}
\newlength{\pdotdiam}
\NewDocumentCommand{\mytikz}{O{-0.6ex}}{\tikz[baseline=#1,radius=\gs]}

\def\psbox{ \draw[fill=white] (-3pt, -3pt) rectangle (3pt,3pt) ; }

\tikzset{
pdot/.style={draw, black, circle, fill, minimum size=\pdotdiam, inner sep=0},
dline/.style={draw, black, line cap=round, thick},
pline/.style={draw, black, line cap=round, thick, dash pattern=on 2pt off 2pt},
psline/.style={draw, black, line cap=round, thick, postaction={ decorate, decoration={markings, mark=at position 0.5 with {\psbox}} } },
}

\NewDocumentCommand \nl{m}
{
 node[below,align=center] {$#1$}
}

\NewDocumentCommand \pdot{O{}}{
\coordinate[pdot] (p1) at (0,0);
\node[below] at (p1) {$#1$};
}

\NewDocumentCommand \hortwopattern{O{}O{}}{
\coordinate[pdot] (p1) at (0,0);
\node [below] at (p1) {$#1$};
\coordinate[pdot] (p2) at (\gs,0);
\node [below] at (p2) {$#2$};
}

\NewDocumentCommand \vertwopattern{O{}O{}}{
\coordinate[pdot] (p1) at (0,0);
\node [below] at (p1) {$#1$};
\coordinate[pdot] (p2) at (0,\gs);
\node [above] at (p2) {$#2$};
}

\NewDocumentCommand \threepattern{O{}O{}O{}}{
\coordinate[pdot] (p1) at (0,0);
\node [below] at (p1) {$#1$};
\coordinate[pdot] (p2) at (\gs,0);
\node [below] at (p2) {$#2$};
\coordinate[pdot] (p3) at (0.5\gs,\gs);
\node [above] at (p3) {$#3$};
}

\NewDocumentCommand \ddline{mmo}{
  \IfNoValueTF{#3}
  {\draw[dline] (#1) -- (#2);}
  {\draw[dline] (#1) .. controls ($($(#1)!0.25!(#2)$)!#3!90:(#2)$) and ($($(#1)!0.75!(#2)$)!#3!90:(#2)$) .. (#2);}
}

\NewDocumentCommand \ppline{mmo}{
  \IfNoValueTF{#3}
  {\draw[pline] (#1) -- (#2) ; }
  {\draw[pline] (#1) .. controls ($($(#1)!0.25!(#2)$)!#3!90:(#2)$) and ($($(#1)!0.75!(#2)$)!#3!90:(#2)$) .. (#2);}
}

\NewDocumentCommand \dpline{mmo}{
  \IfNoValueTF{#3}
  { \draw[dline] (#1) -- ($(#1)!0.45!(#2)$) ;
    \draw[pline] ($(#1)!0.45!(#2)$) -- (#2) ;
  }
  {
    \begin{scope}
     \clip (#1) rectangle ($($(#1)!0.5!(#2)$)!#3!90:(#2)$);    
     \draw[dline] (#1) .. controls ($($(#1)!0.25!(#2)$)!#3!90:(#2)$) and ($($(#1)!0.75!(#2)$)!#3!90:(#2)$) .. (#2);
    \end{scope}
    \begin{scope}
     \clip (#2) rectangle ($($(#1)!0.5!(#2)$)!#3!90:(#2)$);
     \draw[pline] (#1) .. controls ($($(#1)!0.25!(#2)$)!#3!90:(#2)$) and ($($(#1)!0.75!(#2)$)!#3!90:(#2)$) .. (#2);
    \end{scope}
  }
}

\begin{document}

\title{\boldmath A first comparison of Kinetic Field Theory with Eulerian Standard Perturbation Theory}

\author[a,1]{Elena Kozlikin,\note{Corresponding author.}}
\author[b]{Robert Lilow,}
\author[c]{Felix Fabis}
\author[a]{and Matthias Bartelmann}


\affiliation[a]{Heidelberg University, Institut f\"ur Theoretische Physik, Philosophenweg 16, 69120 Heidelberg, Germany}
\affiliation[b]{Department of Physics, Technion, Haifa 3200003, Israel}
\affiliation[c]{Heidelberg University, Zentrum f\"ur Astronomie, Institut f\"ur Theoretische Astrophysik, Philosophenweg 12, 69120 Heidelberg, Germany}

\emailAdd{elena.kozlikin@uni-heidelberg.de}
\emailAdd{rlilow@campus.technion.ac.il}
\emailAdd{felix.fabis@posteo.de}
\emailAdd{bartelmann@uni-heidelberg.de}

\abstract{
We present a detailed comparison of the newly developed particle-based Kinetic Field Theory framework for cosmic large-scale structure formation with the established formalism of Eulerian Standard Perturbation Theory. We highlight the qualitative differences of both approaches by a comparative analysis of the respective equations of motion and implementation of initial conditions. A natural starting point for a first quantitative comparison is given by the non-interacting regime of free-streaming kinematics. 
Our results suggest that Kinetic Field Theory contains a complete resummation of Standard Perturbation Theory in this regime. 
We further show that the exact free-streaming solution of Kinetic Field Theory cannot be recovered in any finite order of Standard Perturbation Theory. Kinetic Field Theory
should therefore provide a better starting point for perturbative treatments of non-linear structure formation.}


\maketitle
\flushbottom

\section{Introduction} \label{sec01}

Over the past two decades, the analytical investigation of cosmic large scale structure formation in cold dark matter (CDM) has attracted considerable interest. Understanding the evolution of density fluctuations in CDM under the influence of gravity is crucial in order to analyse data on large-scale structure and to constrain cosmological models from future surveys such as Euclid, LSST, eBOSS and DESI \cite{Euclid, LSST, eBOSS, DESI}. While numerical $N$-body simulations provide excellent results, they remain computationally costly. To provide the required accuracy of results at scales in the non-linear regime simulations require large volumes and many realisations in order to reduce sample variance. Suitable analytical techniques however can offer a faster and more flexible means of exploring the parameter space of non-standard cosmological models.

Analytical techniques can also help to understand the results of complex simulations in simple analytical terms. Starting with a seminal paper by Crocce \& Scoccimaro \cite{Crocce2006}, numerous concepts originating in field theory like resummation, the renormalisation group \cite{Matarrese, Pietroni2008}, large-$N$ expansion \cite{Valageas2004} or effective field theories \cite{Pietroni2012,Hertzberg2014,Konstandin2019} have been employed to extend (Eulerian) Standard Perturbation Theory (SPT) into the non-linear regime. While this has substantially improved predictive power on the scales of baryonic-acoustic oscillations (BAO), smaller scales beyond wavenumbers of $k = 1 \, h\,\mathrm{Mpc}^{-1}$ are still outside the reach of these methods. The ability of large and costly $N$-body simulations to describe these small scales is based on the fact that they follow the phase-space evolution of individual particles. Since trajectories do not cross in phase-space there is no need for approximations like the single-stream approximation which fundamentally limit the accuracy of analytical schemes based on Eulerian fluid dynamics \cite{Pietroni2012}.

In contrast to the path integral formulation of \cite{Matarrese, Pietroni2008} based on fluid dynamics, the Kinetic Field Theory (KFT) approach presented in \cite{Bartelmann2016, Bartelmann2017, bartelmann_cosmic_2019} is based on particle dynamics. This novel approach originates from a field theoretic description of kinetic theory by Das \& Mazenko \cite{Das2012,Das2013} which we adapted to the setting of cosmic structure formation. Using the path integral approach for classical mechanics (cf.~\cite{Martin1973,Gozzi1989,Penco2006}), it directly encodes the Hamiltonian phase-space dynamics and initial statistics of a particle ensemble into a generating functional. Macroscopic quantities, such as the density-fluctuation power spectrum, are then collectively constructed from the microscopic phase-space information at times of interest.

Mimicking the logic of $N$-body simulations, this approach should by construction be able to overcome the limitations introduced by the SPT approximations and allow us to probe scales where the idea of a unique smooth velocity field, \ie the single-stream approximation, breaks down. Simple first-order perturbation theory \cite{Bartelmann2016} relative to modified Zel'dovich trajectories \cite{Bartelmann2015a}, free-streaming with artificially removed damping \cite{Bartelmann2017} and non-perturbative mean-field approximations for the interaction \cite{bartelmann2020} have shown remarkable similarity with $N$-body results down to scales much smaller than those accessible with SPT.

In this paper we provide the first detailed comparison of KFT with the established SPT, prompted by the encouraging results obtained with KFT. We start in \autoref{sec02} with a brief recapitulation of the necessary KFT basics and then investigate the implementation of initial statistics. As shown in \cite{Bartelmann2016}, due to the particle-based nature of KFT, one needs to sample initial macroscopic information into an initial particle phase-space distribution. We present a more generalised derivation of this step that facilitates the comparison with SPT.

In \autoref{sec03} we then go through the important steps in the derivation of the equations of motion of SPT from the kinetic equations equivalent to KFT. While this calculation is well-known, it will allow us to precisely point out various approximations that are typically made and why they are absent in KFT. Along the way we will also clarify the different notions of averaging in KFT and SPT, and how corresponding macroscopic fields in both frameworks should be identified. We will see that if SPT was formulated in terms of the macroscopic fields that follow `naturally' from the underlying Vlasov equation, the notion of non-linear perturbation theory would be very similar in SPT and KFT. It is only after a change of variables that the physical meaning of using perturbative solutions of the respective equations of motion becomes drastically different between the two theories. We will then show that the natural starting point for quantitatively studying the inherent difference between SPT and KFT is the free-streaming regime.

In \autoref{sec04} we compare the cosmic density power spectrum and bispectrum in the free-streaming regime of both theories. We have already shown in \cite{Fabis2018} that using KFT we can derive these exactly. By expanding the KFT results up to quadratic order in the initial density-fluctuation power spectrum we obtain the same results as in SPT. This suggests that in the free-streaming regime KFT contains a complete resummation of the (free-streaming) SPT perturbation series, making it a more appropriate starting point for perturbative treatments of structure formation. We finally give our conclusions in \autoref{sec05}. 

\subsection{Notation} \label{sec01-01}

We will need to consider the microscopic phase-space coordinates of a large set of $N$ particles confined to a volume $V$. Individual particles are enumerated with \textit{particle labels} $j=1,\ldots,N$ and have positions and momenta denoted as $d$-dimensional vectors $\vq_j$ and $\vp_j$. We combine them into a $2d$-dimensional phase-space coordinate vector as
\begin{equation}
 \vx_j = \cvector{\vq_j \\ \vp_j} \quad \forall j \in \{1,\dots,N\} \;.
\label{eq:phase_space_coordinates}
\end{equation}
For all $N$ particles we bundle these with the help of the tensor product,
\begin{equation}
 \tq = \vq_j \otimes \ve_j \;, \qquad \tp = \vp_j \otimes \ve_j \;, \qquad \tx = \vx_j \otimes \ve_j \;,
\label{eq:phase_space_coordinates_bundled}
\end{equation}
where $\ve_j$ is the canonical base vector in $N$ dimensions with entries $(\ve_j)_i = \delta_{ij}$. The Einstein summation convention is implied unless explicitly stated otherwise or obvious from context. These bold vectors follow the rules of matrix multiplication, inducing a scalar product 
\begin{equation}
 \tens{a} \cdot \tens{b} = \tens{a}^{\top} \, \tens{b} = \sum_{j=1}^N \vec{a}^{\,\top}_j \, \vec{b}_j = \vec{a}_j \cdot \vec{b}_j \;.
\label{eq:bold_scalar_product}
\end{equation}
If $\tens{a},\tens{b}$ are functions of time, we extend the scalar product to additionally imply integration over their time argument,
\begin{equation}
 \tens{a} \cdot \tens{b} = \bsi{t}{t_\mi}{t_\mf}\tens{a}(t) \cdot \tens{b}(t) \;.
\label{eq:bold_dot_product}
\end{equation}
We denote integration over phase-space and the Fourier conjugate to the position $q$ by
\begin{equation}
\begin{split}
 \int_x &\coloneqq \umi{2d}{x} = \umi{d}{q} \umi{d}{p} = \int_q \int_p \;, \\
 \int_h &\coloneqq \fmi{d}{h} \;.
\end{split}
\label{eq:integral_shorthandles}
\end{equation}

\section{Kinetic Field Theory} \label{sec02}

\subsection{Generating functional} \label{sec02-01}
As described in detail in \cite{Bartelmann2016}, KFT encodes both the dynamics and the initial statistics of a non-equilibrium many-particle system in a generating functional. Time evolution is encoded by a functional integral over all possible phase-space trajectories of the individual particles. Since we are dealing with classical particles, a functional delta distribution is introduced that assigns a non-vanishing weight only to those trajectories obeying the classical equations of motion. The stochastic element is introduced by averaging over an initial phase-space probability distribution $\mathcal{P}(\txi)$. The generating functional is thus given by
\begin{equation}
 Z = \usi{\txi} \mathcal{P}(\txi) \bpint{\tx(t)}{\txi}{} \, \dirac\bigl[ \tens{E}[\tx(t)] \bigr] \;,
\label{eq:KFT_GF_definition}
\end{equation}
where $\tens{E}[\tx(t)] = 0$ is the equation of motion for the $N$-particle system. We describe phase-space dynamics by Hamilton's equations, assumed to be of the form
\begin{equation}
 \tens{E}(\tx) = \left(\partial_t + \tens{F}\right) \tx + \tens{\nabla}_q \, V = 0 \;, \quad \tens{F} = F_j \otimes \vec{e}_j \;, \quad F_j \vec{x}_j = \cvector{ -\nabla_{p_j} \\ \nabla_{q_j} } \mathcal{H}_0 \;.
\label{eq:KFT_EoM}
\end{equation}
The matrix $F_j$ encodes the linear equation of motion for individual particles, $V(\vq,t)$ is the interaction potential and $\mathcal{H}_0$ denotes the free part of the Hamiltonian. We restrict ourselves to systems of $N$ identical particles in the absence of external forces and assume that the force between particles acts instantaneously and only depends on their configuration-space positions. We can then write $V$ as a superposition of $N$ single particle potentials $v$. Following \cite{Bartelmann2016}, the generating functional can then be written as
\begin{equation}
  Z = \IC \bpint{\tx(t)}{\txi}{} \upi{\tchi(t)} \e^{ \mi \left( \tchi \cdot \left(\partial_t + \tens{F}\right) \tx + \Phi \cdot \sigma \cdot \Phi \right) } \;.
\label{eq:KFT_GF_explicit}
\end{equation}
In \eqref{eq:KFT_GF_explicit} we have expressed the functional delta distribution from \eqref{eq:KFT_GF_definition} as a Fourier transform, which introduces the auxiliary field $\tchi(t)$ as conjugate to the phase-space coordinate $\tx(t)$. We have furthermore introduced the tuple
\begin{equation}
 \Phi = \cvector{ \Phi_f \\ \Phi_B }
\label{eq:KFT_fields_potential}
\end{equation}
that contains the two collective fields $\Phi_f$ and $\Phi_B$. For tuples $\Phi(r)$ the dot product implies contraction over component indices and integration over general phase-space arguments $r = (t_r,\vx_r)$. The integral over the initial phase-space distribution was abbreviated as $\mathrm{d}\Gamma_\mi = \mathrm{d}\txi \, \mathcal{P}(\txi)$.

The Klimontovich phase-space density
\begin{equation}
  \Phi_f(\vx,t) = \sum_{j=1}^N \dirac\bigl(\vx - \vx_j(t)\bigr) = \sum_{j=1}^N  \dirac\bigl(\vq - \vq_j(t)\bigr) \, \dirac\bigl(\vp - \vp_j(t)\bigr)
\label{eq:KFT_phase_space_density}
\end{equation}
contains the information which particles of the ensemble occupy a phase-space state $\vx$ at time $t$. The so-called `response field' 
\begin{equation}
   \Phi_B(\vx,t) = \sum_{j=1}^N \vchi_{p_j}(t) \cdot \nabla_q \dirac\bigl(\vq - \vq_j(t)\bigr) \, \dirac\bigl(\vp - \vp_j(t)\bigr)
\label{eq:KFT_response_field}
\end{equation}
encodes the deviation of all individual particles from their inertial trajectories as a linear response to some disturbance, which in our case is caused by the interaction with all other particles. For a compact notation we have defined the `interaction matrix'
\begin{equation}
    \sigma(r,r') = - \frac{1}{2} \, \dirac(t_r - t_{r'}) \, v(\vq_r, \vq_{r'}) \matrix{cc}{ 0 & 1 \\ 1 & 0} \;,
\end{equation}
containing the single-particle potential $v$.

Next we introduce source fields $H_f$ and $H_B$ for the collective fields $\Phi_f$ and $\Phi_B$, as well as source fields $\tJ$ and $\tK$ coupling to $\tx$ and $\tchi$, respectively. Replacing
\begin{equation}
 \vx_j(t) \rightarrow \hat{\vx}_j(t) \coloneqq \fd{}{\vec{J}_j(t)} \quad \textrm{and} \quad \vchi_j(t) \rightarrow \vochi_j(t) \coloneqq \fd{}{\vec{K}_j(t)} 
\label{eq:KFT_particle_operators}
\end{equation}
in the collective fields $\Phi$ turns them into operators $\hat{\Phi}$. This allows us to rewrite the generating functional as
\begin{equation}
\begin{split}
  Z[H,\tJ,\tK] &= \e^{\mi \, H \cdot \hat{\Phi}} \, \e^{\mi \, \hat{\Phi} \cdot \sigma \cdot \hat{\Phi} } \IC \bpint{\tx}{\txi}{} \upi{\tchi} \e^{ \mi \left( \tchi \cdot \left(\partial_t + \tens{F}\right) \tx + \tJ \cdot \tx + \tK \cdot \tchi \right) } \\
  &= \e^{\mi \, H \cdot \hat{\Phi}} \, \e^{\mi \, \hat{\Phi} \cdot \sigma \cdot \hat{\Phi} } \, Z_{0}[\tJ,\tK] \;.
\end{split}
\label{eq:KFT_GF_operator_version}
\end{equation}
Cumulants (\ie the connected part of correlation functions) of the collective fields can then be obtained by taking appropriate functional derivatives of the logarithm of the generating functional with respect to the source fields,
\begin{equation}
\begin{split}
 G_{\alpha_1 \dots \alpha_n}(1, \dots, n) \coloneqq &\ave{\Phi_{\alpha_1}(1) \dots \Phi_{\alpha_n}(n)}_{\mathrm{connected}} \\
 = &\fd{}{H_{\alpha_1}(1)} \dots \fd{}{H_{\alpha_n}(n)} \, \ln Z[H] \, \bigg|_{H=0} \;.
\end{split}
\label{eq:KFT_cumulants_definition}
\end{equation}
As shown in \cite{Bartelmann2016}, the path integrals in the \emph{free} generating functional $Z_{0}[\tJ,\tK]$ can be performed once the Green's function $\mathcal{G}$ of the free equation of motion of a single particle is known. One finds
\begin{equation}
 Z_{0}[\tJ,\tK] = \usi{\txi} \mathcal{P}(\txi) \, \e^{ \mi \, \tens{J} \cdot \bar{\tx} } \;,
\label{eq:KFT_GF_free_solution}
\end{equation}
where the solution to the equations of motion $\bar{\tx}(t)$ is defined as
\begin{equation}
 \bar{\tx}(t) = \tens{\mathcal{G}}(t,t_\mi) \, \txi - \bsi{t'}{t_\mi}{t_\mf} \tens{\mathcal{G}}(t,t') \tK(t') \;,
\label{eq:KFT_free_trajectories}
\end{equation}
and the linear particle propagator is of the general form
\begin{equation}
 \tens{\mathcal{G}}(t,t') = \mathcal{G}(t,t') \otimes \mathcal{I}_N \quad \textrm{with} \quad \mathcal{G}(t,t') = \matrix{cc}{ \gqq(t,t') \, \mathcal{I}_d & \;\; \gqp(t,t') \, \mathcal{I}_d \\ \gpq(t,t') \, \mathcal{I}_d & \;\; \gpp(t,t') \, \mathcal{I}_d } \;,
\label{eq:KFT_particle_propagator}
\end{equation}
with $\mathcal{I}_d$ denoting the $d$-dimensional identity matrix. The additional source term $\tK$ in \eqref{eq:KFT_free_trajectories} allows us to push particles away from their inertial motion if the particles are interacting via an interaction potential. For free-streaming particles we can thus set $\tK=0$. 

We can now understand what perturbation theory in KFT physically means: Expanding the exponential interaction operator in \eqref{eq:KFT_GF_operator_version} to first order in the interaction matrix $\sigma$ evaluates the force exerted on a particle along its free trajectory due to the interactions with all other particles on their respective free trajectories. Since the auxiliary fields $\vchi_j$ in the response field \eqref{eq:KFT_response_field} are replaced by functional operators with respect to $\vec{K}_j$, this force is then inserted for $\vec{K}_j$ in \eqref{eq:KFT_free_trajectories} and modifies the free trajectory of that particle. The actual perturbed physical quantity in the perturbation theory in KFT are therefore the trajectories of particles and not the interaction potential. It is important to keep this fact in mind to understand the difference in physical pictures between perturbation theory in KFT and SPT later.

\subsection{Averages over initial statistics} \label{sec02-02}

In the case of cosmic structure formation the only information we have about the initial state of our system is of a statistical nature. Usually, we only know the low $n$-point order correlations of the cosmic density and velocity fields from observations. This information must be represented by our $N$-particle ensemble. The appropriate phase-space distribution $\mathcal{P}(\txi)$ can be constructed in the following two steps.

As a first step, we must be able to describe a general realisation of the smooth initial macroscopic particle number density and momentum fields, $\ini{\rho}(\vq)$ and $\vini{P}(\vq)$, as an average over the ensemble of $N$-particle systems. We thus need to find the appropriate sampling distribution $\mathcal{P}_{\mS}$ conditioned such that the \emph{sampling average} $\ave{\dots}_{\mS}$ of the collective number and momentum density fields,
\begin{alignat}{2}
  \Phi_\rho(\vq,t) &= \int_{p} \Phi_f(\vx,t)& &= \sum_{j=1}^N \dirac\bigl(\vq - \vq_j(t)\bigr) \;,
  \label{eq:KFT_number_density} \\
  \Phi_{\vec{\Pi}}(\vq,t) &= \int_{p} \vp \, \Phi_f(\vx,t)& &= \sum_{j=1}^N \vp_j(t) \, \dirac\bigl(\vq - \vq_j(t)\bigr) \;,
  \label{eq:KFT_momentum_density}
\end{alignat}
reproduces the given initial macroscopic field values,
\begin{equation}
 \ave{\Phi_\rho(\vq,t_\mi)}_{\mS} \stackrel{!}{=} \ini{\rho}(\vq) \;, \qquad \ave{\Phi_{\vec{\Pi}}(\vq,t_\mi)}_{\mS} \stackrel{!}{=} \vini{\Pi}(\vq) = \ini{\rho}(\vq) \, \vini{P}(\vq) \;.
\label{eq:KFT_sampling_constraints}
\end{equation}
For point-like particles a very simple way to satisfy these constraints is to use Poisson sampling, where particles $j$ are placed randomly at positions $\vqi_j$ inside the volume $V$ with a probability density proportional to the macroscopic number density at that point. They are furthermore assigned a momentum corresponding to the value of the macroscopic momentum field. This sampling distribution reads
\begin{equation}
 \mathcal{P}_{\mS}\bigl(\txi \, \big| \, \tens{Y}^{(\mi)} \bigr) = \prod_{j=1}^N \, \frac{1}{N} \, \ini{\rho}_j \, \dirac\bigl(\vpi_j - \vini{P}_j\bigr) 
 \quad \textrm{with} \quad \tens{Y}^{\,(\mi)} = \cvector{\ini{\rho}_j \\ \vini{P}_j} \otimes \ve_j \;, 
\label{eq:KFT_conditioned_sampling_distribution}
\end{equation}
where $\ini{\rho}_j = \ini{\rho}(\vqi_j)$ and likewise for $\vini{P}$, \ie $\tYi$ bundles the initial field values at the starting positions of the sampling particles. We emphasise that this choice of sampling does not induce any connected correlations; there is, for example, no dispersion of particle momenta around the macroscopic flow $\vini{P}$.

In the second step we consider $\tYi$ to be a random field with multi-point probability density $\mathcal{P}_{\mF}$ fixed by the $n$-point correlations of the density and momentum fields. Averages over this distribution of the fields will be denoted as $\ave{\dots}_{\mF}$. We can therefore write the expectation value for a time-evolved observable $A(\tx(t))$ averaged over the initial field values as
\begin{align}
 \ave{A\bigl(\tx(t')\bigr)} &= \ave{ \ave{A\bigl(\tx(t')\bigr)}_{\mS} }_{\mF}
 \label{eq:KFT_general_expectation_value} \\
 &= \usi{\tYi} \mathcal{P}_{\mF}\bigl(\tYi\bigr) \usi{\txi} \mathcal{P}_{\mS}\bigl(\txi \, \big| \, \tYi \bigr) \bpint{\tx(t)}{\txi}{} \, A\bigl(\tx(t')\bigr) \dirac\bigl[ \tens{E}(\tx(t)) \bigr] \nonumber \\
 &= \int \mathrm{d}\txi \usi{\tYi} \mathcal{P}_{\mF}\bigl(\tYi\bigr) \, \mathcal{P}_{\mS}\bigl(\txi \, \big| \, \tYi \bigr) \bpint{\tx(t)}{\txi}{} \, A\bigl(\tx(t')\bigr) \dirac\bigl[ \tens{E}(\tx(t)) \bigr] \;, \nonumber
\end{align}
allowing us to identify the initial phase-space distribution $\mathcal{P}(\txi)$ in \eqref{eq:KFT_GF_definition} as
\begin{equation}
 \mathcal{P}\bigl( \txi \bigr) = \usi{\tYi} \mathcal{P}_{\mF}\bigl(\tYi\bigr) \, \mathcal{P}_{\mS}\bigl(\txi \, \big| \, \tYi \bigr) \;.
\label{eq:KFT_initial_phase_space_distribution}
\end{equation}

The second line of \eqref{eq:KFT_general_expectation_value} encapsulates the general idea of how KFT obtains time-evolved expectation values of collective observables $A$. First, one solves the evolution of an $N$-particle system starting from $\txi$ in an abstract sense. The actual calculations are performed, in the simplest case, as a perturbation series in the interaction potential. The result is used to calculate the value of $A(\tx)$ in terms of $\txi$. This observable is then sampling-averaged with the result being expressed in terms of $\tYi$. We will see shortly that in SPT this process is represented by writing down and solving evolution equations for the macroscopic observables $A$ themselves. The final step is to average over the initial macroscopic random fields. In SPT this would amount to averaging a (perturbative) solution $A(t;A(t_\mi))$ over $A(t_\mi)$. 

In the formalism of KFT, we use that for any observable that can be directly expressed in terms of $\tx(t)$, we can exchange the order of taking the averages in the third line of \eqref{eq:KFT_general_expectation_value}. We thus integrate over the field values $\tYi$ before integrating over the microscopic $\txi$. This allows us to imprint the initial correlations of the macroscopic fields directly onto the statistics of the microscopic particle ensemble.

In cosmic structure formation the distribution $\mathcal{P}_{\mF}$ is chosen as a statistically homogeneous and isotropic Gaussian. For the explicit calculation of $\mathcal{P}(\txi)$ we refer the reader to \cite{Bartelmann2016}. Its exact treatment when calculating free-streaming cumulants can be found in \cite{Fabis2018}.

\section{Standard Perturbation Theory} \label{sec03}

\subsection{The Vlasov equation} \label{sec03-01}

There is a fundamental difference between KFT and SPT in how they describe the dynamics of a system. While KFT operates on the microscopic phase-space of particles, SPT gives up part of this information and works with equations of motion for the macroscopic observables that one is ultimately interested in, \eg density and momentum fields. In order to show which approximations separate the two approaches we will start with the Klimontovich equation, describing the exact dynamics of the phase-space density \eqref{eq:KFT_phase_space_density}, as the basis for KFT as well as SPT. The Klimontovich equation is given by
\begin{equation}
 \partial_t \Phi_f(\vx,t) + \frac{\vp}{m} \cdot \nabla_q \Phi_f(\vx,t) - \int_{x'} \left( \nabla_q v(\vq,\vq^{\,\prime}) \, \Phi_f(\vx^{\,\prime},t) \right) \cdot \nabla_p \Phi_f(\vx,t) = 0 \;,
\label{eq:SPT_klimontovich_equation}
\end{equation}
where we assumed that the Hamiltonian equations of motion have the form
\begin{equation}
 \tder{\vq_j}{t}= \frac{\vp_j}{m} \;, \qquad \tder{\vp_j}{t} = -\nabla_{q_j} \sum_{i \neq j} v(\vq_j, \vq_i) = - \int_x \, \nabla_{q_j} v(\vq_j,\vq) \, \Phi_f(\vx,t) 
\label{eq:SPT_particle_EoM}
\end{equation}
and self-interaction of particles is suppressed by hand, \ie $\nabla_q v(\vq,\vq^{\,\prime}) \big|_{\vq^{\,\prime} = \vq} \equiv 0$. We stress that \eqref{eq:SPT_klimontovich_equation} encodes the same dynamical information as the generating functional of KFT. Perturbation theory in KFT is formally equivalent to solving the linearised version of \eqref{eq:SPT_klimontovich_equation} and then iteratively inserting the obtained solution into the quadratic term to generate perturbative corrections in orders of the interaction potential $v$. The technical difference to working at the level of \eqref{eq:SPT_klimontovich_equation} is that KFT directly perturbs the Hamiltonian equations of motion of the individual particles and
builds up time-evolved collective information from the perturbed trajectories.

There are different notions of how the transition from the spiky distribution $\Phi_f$ to a smooth one can be made. A very common direct approach \cite{Buchert2005, Pietroni2012}
is to work with a single instance of the $N$-particle system and implement explicit local coarse-graining in phase-space
\begin{equation}
 \bar{\Phi}_f(\vq, \vp, t) \coloneqq \int_{\Delta q} \int_{\Delta p} \mathcal{W}_{\ell_q}(\Delta\vec{q}) \, \mathcal{W}_{\ell_p}(\Delta\vec{p}) \, \Phi_f(\vq + \Delta\vq,\vp + \Delta\vp, t)\, ,
\label{eq:spatial_coarse_graining}
\end{equation}
using normalised window functions $\mathcal{W}_{\ell_q}, \mathcal{W}_{\ell_p}$ with coarse-graining scales $\ell_q,\ell_p$ for position and momentum, respectively.

In our case the coarse graining process is already implicitly contained in the ensemble average over initial conditions. As pointed out in \cite{Ichimaru} there is no dynamical interaction between ensemble members, so any distribution of the full microscopic phase-space state $\tx$ is conserved along the phase-space trajectories due to Liouville's theorem. Taking the ensemble average at any later time $t > t_\mi$ can thus be directly mapped to the ensemble average over initial conditions. The explicit coarse-graining scales $\ell_q, \ell_p$ are replaced by the implicit coarse-graining scales of the initial macroscopic fields $\ini{\rho},\vini{P}$.

As pointed out in \cite{Buchert2005}, the two notions of coarse-graining are mathematically similar when one re-interprets the window functions $\mathcal{W}$ as a probability distribution in the sense of $\mathcal{P}_\mS$. The dynamical equations of the coarse-grained sector will in fact have the same form. Let us now define
\begin{align}
 \ave{ \Phi_f(\vx_1,t) }_{\mS} &\eqqcolon \psd{1}(\vx_1,t)  \label{eq:SPT_one_point_phase_space_density} \;, \\
 \ave{ \Phi_f(\vx_1,t) \, \Phi_f(\vx_2,t) }_{\mS} &\eqqcolon \dirac\left(\vx_1 - \vx_2\right) \psd{1}(\vx_1,t) + \psd{2}(\vx_1,\vx_2,t) \;, \label{eq:SPT_two_point_phase_space_density}
\end{align}
where $\psd{1}, \psd{2}$ are the one- and two-particle distribution functions familiar from the BBGKY hierarchy. The first term of \eqref{eq:SPT_two_point_phase_space_density} is shot-noise arising from the discrete particle nature of $\Phi_f$. We may further separate the two-particle function as
\begin{equation}
 \psd{2}(\vx_1,\vx_2,t) = \psd{1}(\vx_1,t) \, \psd{1}(\vx_2,t) + \psd{2}_{\mC}(\vx_1,\vx_2,t) \;,
\label{eq:SPT_two_point_phase_space_density_connected}
\end{equation}
where the second term contains connected correlations between particles. Next we define a mean, macroscopic potential
\begin{equation}
 \phi(\vq,t) \coloneqq \int_{x'} v(\vq,\vq^{\,\prime}) \, \psd{1}(\vx^{\,\prime},t) \;.
\label{eq:SPT_ensemble_averaged_potential}
\end{equation}
With these definitions the ensemble average of \eqref{eq:SPT_klimontovich_equation} reads
\begin{equation}
\begin{split}
 \partial_t \psd{1}(\vx,t) &+ \frac{\vp}{m} \cdot \nabla_q \psd{1}(\vx,t) - \bigl( \nabla_q \phi(\vq,t) \bigr) \cdot \nabla_p \psd{1}(\vx,t) \\
 &= \int_{x'} \bigl( \nabla_q v(\vq,\vq^{\,\prime}) \bigr) \cdot \nabla_p  \psd{2}_{\mC}(\vx,\vx^{\,\prime},t) 
\end{split}
\label{eq:SPT_one_point_BBGKY}
\end{equation}
where the shot noise contribution of \eqref{eq:SPT_two_point_phase_space_density} drops out because particles do not interact with themselves. The left-hand side of this equation describes the system above the coarse-graining scale while the right hand side has to account for any effects below it like small-scale collisions.

According to the BBGKY hierarchy, $\psd{2}_{\mC}$ will obey a dynamical equation sourced by $\psd{3}$, which in turn is sourced by higher order correlation functions and so on. In the context of cosmic structure formation there are various approaches to treating the small-scale fluctuations in order to close the hierarchy \cite{Buchert2005,Pietroni2012,Ma2004}. We will restrict ourselves to the simple \textit{dust} model by setting $\psd{2}_{\mC} = 0$ which leads to the standard form of the SPT equations. This assumes that small-scale collisions are negligible compared to the large-scale gravitational forces described by the mean potential $\phi$, \ie treating CDM as a \emph{collisionless} system. Since $v$ is the single-particle gravitational potential, \eqref{eq:SPT_ensemble_averaged_potential} is replaced by Poisson's equation and we arrive at the Vlasov-Poisson system of equations,
\begin{align}
 \partial_t \psd{1}(\vx,t) + \frac{\vp}{m} \cdot \nabla_q \psd{1}(\vx,t) - \bigl( \nabla_q \phi(\vq,t) \bigr) \cdot \nabla_p \psd{1}(\vx,t) &= 0 \;,\label{eq:SPT_vlasov_equation} \\
 \nabla_q^2 \phi(\vq,t) &= 4\pi G m^2 \int_{p} \psd{1}(\vx,t) \;.\label{eq:SPT_poisson_equation}
\end{align}

We want to stress here that because KFT performs the perturbative expansion of its dynamics at the exact level of \eqref{eq:SPT_klimontovich_equation} and only performs the ensemble average afterwards, no assumptions about $\psd{2}_{\mC}$ or any higher-order correlation functions need to be made. Their effects are successively taken into account with rising orders of perturbation theory in the interaction \cite{Viermann2015}. While our particular choice of sampling average will set $\psd{2}_{\mC} = 0$ at the initial time due to the Poisson sampling process, $\psd{2}_{\mC}$ will be generated dynamically by particle interactions. Its dynamical equation is sourced by a term of the form $ \psd{1} \nabla_q v \cdot \nabla_p \psd{1}$ \cite{Ichimaru}, which one can easily relate to a two-particle term in the first order of the KFT perturbation series for the phase-space density cumulant $G_{ff}(1,2)$.

However, since for a system with fixed mean mass density the potential must scale as $v \propto 1/\mpd \propto 1/N$, this particular two-particle term is usually discarded as subdominant shot-noise in the thermodynamic limit of $N \to \infty$. The same then holds for any terms of higher order in the KFT perturbation series which can be linked to terms in the BBGKY hierarchy that include $\psd{2}_{\mC}$. Analogous arguments hold for any $\psd{n}_\mC$ with $n > 2$. Discarding shot-noise terms is thus equivalent to treating CDM as a collisionless gas only subject to the effects of its own coarse-grained gravitational potential.

This confirms that the assumption $\psd{2}_{\mC} = 0$ of SPT is in fact unproblematic as long as the constituents are microscopic and thus light enough to ensure large mean particle densities on scales of interest in the range of $\mathrm{Mpc}$ \cite{Bernardeau2002}. However, in applications beyond cosmology (see, for instance, \cite{bartelmann_cosmic_2019} where KFT is applied to a system of Rydberg-atoms) such discreteness effects may become important. Also, if the sampling average contains connected correlations, for example when the particles used to sample the initial density field can no longer be considered point-like, then the $\psd{n}_\mC$ will contribute not only on the shot-noise level to the statistical evolution. The KFT framework captures both these cases by construction.

\subsection{Hierarchy of moments} \label{sec03-02}

The system of partial differential equations \eqref{eq:SPT_vlasov_equation} and \eqref{eq:SPT_poisson_equation} is still a very hard system to solve. Therefore one introduces the moments
\begin{alignat}{3}
 \rho(\vq,t) &\coloneqq \int_{p} \psd{1}(\vx,t) &&= \ave{ \int_{p} \Phi_f(\vx,t) }_{\mS} &&\eqqcolon \bigl\langle \Phi_\rho(\vq,t) \bigr\rangle_{\mS} \label{eq:SPT_particle_density} \;, \\
 \vec{\Pi}(\vq,t) &\coloneqq \int_{p} \vp \, \psd{1}(\vx,t) &&= \ave{ \int_{p} \vp \, \Phi_f(\vx,t) }_{\mS} &&\eqqcolon \ave{ \Phi_{\vec{\Pi}}(\vq,t) }_{\mS} \label{eq:SPT_momentum_density} \;, \\
 T^{ij}(\vq,t) &\coloneqq \int_{p} p^{i} p^{j} \, \psd{1}(\vx,t) &&= \ave{ \int_{p} p^{i} p^{j} \, \Phi_f(\vx,t) }_{\mS} &&\eqqcolon \ave{ \Phi^{ij}_{T}(\vq,t) }_{\mS} \label{eq:SPT_stress_tensor} \;,
\end{alignat}
defining the particle number density $\rho$, the momentum density $\vec{\Pi}$ and the stress tensor $T^{ij}$, respectively, which relate to the sampling averages of microscopic collective fields in KFT as shown above. Taking the zeroth and first moment of the Vlasov equation \eqref{eq:SPT_vlasov_equation} then leads to the following system of equations,
\begin{align}
 \partial_t \, \rho_\mm + \partial_i \, \Pi^i &= 0 \label{eq:SPT_continuity_equation} \;,\\
 \partial_t \, \Pi^{i} + \frac{1}{m} \partial_j \, T^{ij} + \frac{1}{m} \rho_\mm \, \partial^i \phi &= 0 \label{eq:SPT_euler_equation} \;,\\
 \nabla_q^2 \phi &= 4\pi G m \, \rho_m \label{eq:SPT_poisson_equation_with_mass_density} \;,
\end{align}
where we introduced the mass density $\rho_\mm = m \rho$. The first two are evidently the continuity equation and a non-standard form of the Euler equation. Since they are well-known in the context of hydrodynamics they are often called \emph{fluid} equations. In this form they exhibit an interesting property: The only non-linearity enters through the source term $\rho\, \partial^i \phi $, which couples the lower-order moment $\rho$ to the coarse-grained potential $\phi$. Thus, when phrased in terms of these \emph{natural moments} of the phase-space density, the fluid equations derived from the Vlasov equation suggest a perturbative expansion in orders of the potential interaction. However, because each moment is linearly sourced by the next higher one through the divergence term, knowledge of the linear solution means solving an infinite hierarchy of linear coupled equations. The solution of this infinite hierarchy is equivalent to the ensemble-averaged solution of the free motion of all individual particles, which is exactly what KFT uses as the starting point of perturbation theory. In this sense KFT encompasses the entire hierarchy of moments of the Vlasov equation.

\subsection{Velocity field} \label{sec03-03}

In order to truncate this hierarchy one combines the moments into a velocity field,
\begin{equation}
\vec{v}(\vq,t) \coloneqq \frac{\vec{\Pi}(\vq,t)}{\rho_m(\vq,t)} \;,
\label{eq:SPT_velocity_field}
\end{equation}
and the velocity dispersion around this mean flow,
\begin{align}
 \sigma^{ij}(\vq,t) \coloneqq& \, \frac{1}{\rho(\vq,t)} \int_p \left(\frac{p^i}{m} - v^i\right) \left(\frac{p^j}{m} - v^j\right) \psd{1}(\vx,t) \nonumber \\
 =& \, \frac{1}{m\rho_\mm(\vq,t)} T^{ij}(\vq,t) - v^{i}(\vq,t) \, v^{j}(\vq,t) \;. \label{eq:SPT_velocity_dispersion}
\end{align}
The definition of the velocity field as the quotient of two coarse-grained moments makes it a complicated variable to work with in KFT. Perturbation theory in KFT depends on the possibility to express collective fields as operators using \eqref{eq:KFT_particle_operators}. The moments \eqref{eq:SPT_particle_density}--\eqref{eq:SPT_stress_tensor}, which are linear in $\Phi_f$, are thus the most natural choice \cite{Viermann2015}.\footnote{Velocity statistics in KFT are extracted via a velocity-density operator $\hat{\Phi}_{\vec{\Pi}}$ \cite{bartelmann_cosmic_2019}.} Because there is no straightforward expression for an operator that would return the inverse of the density, defining a microscopic velocity field operator $\hat{\Phi}_{\vec{v}}$ is complicated, if not impossible. Even if it did exist, its sampling average would differ from the above macroscopic velocity field \cite{Pietroni2012} since
\begin{equation}
 \bigl\langle\Phi_{\vec{v}}\bigr\rangle_{\mS} = \ave{ \frac{\Phi_{\vec{\Pi}}}{\Phi_\rho} }_{\mS} \neq \frac{ \bigl\langle\Phi_{\vec{\Pi}}\bigr\rangle_{\mS} }{ \ave{\Phi_\rho}_{\mS} } \;.
\label{eq:SPT_no_velocity_field_operator}
\end{equation}
Comparisons between KFT and SPT should therefore be performed in terms of the moments of $\Phi_f$, which are well-defined in both frameworks. In this paper, we restrict ourselves to the comparison of density correlations.

\subsection{The single-stream approximation} \label{sec03-04}

With the help of \eqref{eq:SPT_velocity_field} and \eqref{eq:SPT_velocity_dispersion}, we can now rewrite \eqref{eq:SPT_continuity_equation}--\eqref{eq:SPT_poisson_equation_with_mass_density} as
\begin{align}
 \partial_t \rho_\mm + \partial_i \left(\rho_m \, v^i \right) &= 0 \;, \label{eq:SPT_continuity_equation_with_velocity_field}  \\
 \partial_t v^i + v^j \, \partial_j v^i + \partial^i \tilde{\phi} + \frac{\partial_j \left( \rho_\mm \, \sigma^{ij} \right)}{\rho_\mm }  &= 0 \;, \label{eq:SPT_euler_equation_with_velocity_field} \\
 \nabla_q^2 \tilde{\phi} &= 4\pi G \, \rho_m \;, \label{eq:SPT_poisson_equation_massless_potential} 
\end{align}
where we defined $\tilde{\phi} \coloneqq \phi/m$. In order to close this hierarchy we still need to specify $\sigma^{ij}$. The standard approach in SPT is to neglect all deviations from the mean flow of matter by setting $\sigma^{ij} = 0$, which is why its dynamical model is often called \emph{pressureless dust}.\footnote{See \cite{Buchert2005} for several other analytical ansatzes and their problems.}
It is thus assumed that the system can be described at every point by a single-valued velocity field. This is the so-called \emph{single stream approximation} (SSA). However, this assumption has to break down once macroscopic streams start to cross each other, which severely limits the range of validity of SPT in the context of non-linear cosmic structure formation.

In order to further reduce the number of equations, SPT assumes an irrotational velocity field, $\nabla \times \vec{v} = 0$, which corresponds to a potential flow. This assumption is motivated by the decay of initial vorticity due to the cosmological background expansion in the linear regime. While this is self-consistent in the context of the SSA, a non-vanishing $\sigma^{ij}$ can source vorticity at later times in the evolution of structure \cite{Bernardeau2002}.

We remark here that both the SSA and the potential flow assumption are absent from KFT because the macroscopic velocity field is not part of the dynamical description. Especially with regard to the SSA, it is crucial to note that KFT directly evolves the full phase-space information of all constituent particles of the system, thereby allowing for the crossing of streams in position space without causing any problems. In \cite{Bartelmann2016, bartelmann_cosmic_2019, bartelmann2020} we were able to show that KFT allows us to compute the non-linear density power spectrum far into the non-linear regime up to wave numbers of $k \leq 10 \, h \, \mathrm{Mpc}^{-1}$ with an accuracy typically within $10 \%$ as compared to state of the art fits to numerical results such as \cite{mead_accurate_2015}. This result takes us far beyond the accessible regime of perturbation schemes based on SPT (typically $k < 1 \, h \, \mathrm{Mpc}^{-1}$) and is very likely a direct consequence of multi-streaming being inherently included in KFT. The authors of \cite{mcdonald_large-scale_2018} argue that even though the effects of stream crossing found in numerical simulations appear to be small, it has a self-regulatory property in perturbation theory, such that its effect on the power spectrum obtained from a perturbation theory calculation is large. We should note here that various effective field theory approaches based on Eulerian SPT exist which allow to re-capture at least some part of the physics lost due to the SSA (see for instance \cite{Pietroni2012, Hertzberg2014, Konstandin2019}). These efforts, however, primarily focus on achieving (sub-)percent accuracy in the mildly non-linear regime ($k \leq 0.6 \, h \, \mathrm{Mpc}^{-1}$) for the density power spectrum by employing results from numerical simulations or observations to constrain parameters which are introduced to effectively account for effects of higher-order momenta of the Vlasov-Boltzmann hierarchy.

Although Lagrangian Perturbation Theory (LPT) is technically also free of the SSA, it still breaks down at the epoch of shell-crossing. LPT allows us to study structure formation in terms of the trajectories of fluid elements in configuration-space. The focus of LPT is on perturbing the displacement field, which maps the initial positions of fluid elements onto the final (Eulerian) positions, rather than the density and velocity field. The problem in LPT arises due to the computation of the forces acting on a trajectory in regions where multiple streams meet at the same Eulerian position. 
The total gravitational force acting on a fluid element at this position should be the sum of the contributions from each stream, and the fluid element should then be accelerated by this total force field. Instead, the LPT solutions only accelerate fluid elements with the gravitational force field of their individual stream even after shell-crossing \cite{Buchert1994, Buchert1995}.\footnote{This violates Einstein's equivalence principle of inertial and gravitational mass.} In KFT this issue is avoided by construction: the perturbative expansion is performed in terms of the (gravitational) force acting on each particle. This force is sourced by a density which is given by the superposition of all particle trajectories and thus correctly accounts for multi-streaming. It is possible to include multi-streaming in the Lagrangian framework if, analogously to KFT, forces are sourced by a density which is given by the superposition of all streams, and a perturbative expansion is then performed in terms of the force acting on fluid elements \cite{mcdonald_large-scale_2018}. However, at this point a proper comparison between KFT and this recent approach is not possible, as computations of 3D power spectra for this extension of the Lagrangian formalism do not yet exist to the best of our knowledge.

\subsection{SPT generating functional} \label{sec03-05}

In a last step we take the divergence of \eqref{eq:SPT_euler_equation_with_velocity_field} and insert \eqref{eq:SPT_poisson_equation_massless_potential} to arrive at
\begin{align}
 \partial_t \, \rho_\mm + \rho_\mm \, \theta + v^i \, \partial_i \, \rho_m &= 0 \label{eq:SPT_continuity_equation_with_velocity_divergence} \;, \\
 \partial_t \, \theta + \partial_i \left( v^j \, \partial_j v^i \right) + 4\pi G \rho_m &= 0 \label{eq:SPT_euler_equation_with_velocity_divergence} \;,
\end{align}
where the velocity-divergence field $\theta = \partial_i v^i$ was defined. Introducing the density contrast $\delta = \rho_\mm / \bar{\rho}_\mm - 1$ and Fourier-transforming \eqref{eq:SPT_continuity_equation_with_velocity_divergence} and \eqref{eq:SPT_euler_equation_with_velocity_divergence}, these equations can be completely expressed in terms of $\delta(\vk,t)$ and $\theta(\vk,t)$. Combining them into the field doublet
\begin{equation}
 \varphi \coloneqq \cvector{ \delta \\ - \theta}
\label{eq:SPT_field_doublet}
\end{equation}
and following the same reasoning as in KFT, we can write down a suitable generating functional \cite{Matarrese}
\begin{equation}
 Z_{\textsc{SPT}}[J,K] = \e^{\mi \hat{\chi}_a \gamma_{abc} \hat{\varphi}_b \hat{\varphi}_c} \int \mathcal{D}\varphi_a  \mathcal{D}\chi_a \, \mathcal{P}(\ini{\varphi}_a) \, \e^{\mi \chi_a \left( \partial_t \delta_{ab} + \Omega_{ab} \right) \varphi_b + \mi J_a \varphi_a + \mi K_a \chi_a} \;,
\label{eq:SPT_generating_functional}
\end{equation}
where repeated indices imply summation over field types and integration over arguments. We have, again, introduced an auxiliary field $\chi_a$ conjugate to $\varphi_a$ and appropriate source fields $K_a$ and $J_a$. Their operator versions are defined in analogy to \eqref{eq:KFT_particle_operators}. The initial statistics is contained in $\mathcal{P}(\ini{\varphi}_a)$. The choice of the velocity field as a dynamical variable turns the formerly linear source terms into non-linear ones which are contained in the tensor $\gamma_{abc}$ \cite{Viermann2015}. The gravitational interaction, on the other hand, turns into a linear source term for the velocity encoded by $\Omega_{ab}$, which plays the same role as the force $\tens{F}$ in \eqref{eq:KFT_EoM}.

This drastically alters the physical interpretation of the SPT perturbative expansion in the `interaction operator' $\hat{\chi}_a \gamma_{abc} \hat{\varphi}_b \hat{\varphi}_c$ when compared with perturbative expansion in KFT: The `free' (or linear) theory in SPT describes the independent evolution of Fourier field modes. This already includes gravitational interactions in terms of the coarse-grained potential $\phi$. These field modes are coupled by the non-linear source terms to produce perturbative corrections. In KFT on the other hand, we saw that perturbative corrections are due to the interaction operator $\hat{\Phi} \cdot \sigma \cdot \hat{\Phi}$, while all non-linear effects of free-streaming are exactly described by  the free generating functional \eqref{eq:KFT_GF_free_solution}, which directly uses the non-interacting phase-space evolution of the particle ensemble.

Due to this inherent difference, it is difficult to compare perturbative corrections in both approaches directly. We thus restrict ourselves to the regime of free-streaming particles, where the initial statistics provides us with the same expansion parameter in both, SPT and KFT, as we shall see later. A direct comparison in the gravitationally interacting regime requires the resummation of the interaction in KFT presented in \cite{Lilow_2019} and will be the subject of future work.

\section{Comparison of cosmic power spectra} \label{sec04}

\subsection{Dynamics and initial conditions} \label{sec04-01}

To allow a meaningful comparison of power spectra derived in both KFT and SPT we have to ensure compatibility of the underlying dynamics. In this section we only give the necessary results and provide the more detailed calculations in Appendix \ref{AppendixA}.

The system under consideration is a distribution of a single species of identical cold dark matter particles with mass $m$ following Newtonian dynamics on the expanding background of standard FLRW cosmology, including a cosmological constant. Instead of the scale factor $a(t)$, we choose our time coordinate $\eta$ as
\begin{equation}
 \eta \coloneqq \ln {D_+(t)} \;, \qquad \md\eta = H f \md t \;, \qquad H \coloneqq \frac{\dot{a}}{a} \;, \qquad f \coloneqq \frac{\md \ln D_+}{\md \ln a} \;,
\label{eq:tau_time_definition}
\end{equation}
where $D_+$ is the usual linear growth factor defined as the growing mode solution to
\begin{equation}
 \ddot{\delta} + 2 H \dot{\delta} = \frac{3}{2} \Omega_\mm H^2 \delta \;, \qquad \Omega_\mm = \frac{8 \pi G}{3 H^2} \, \rho_\mm \;.
\label{eq:lin_growth_factor_diff_equation}
\end{equation}
Overdots denote derivatives \wrt time $t$. The growth factor is assumed to be normalised as $D_+(t_\mi) = 1$ at initial time $t_\mi$ and thus $\eta_\mi = 0$. We work in \emph{comoving} coordinates $\vq = \vec{r} / a$ which are at rest relative to the homogeneous expansion of the physical coordinates $\vec{r}$. The momenta $\vp$ of particles and the peculiar velocity field $\vec{u}_{\mathrm{pec}}(\vq,\eta)$ are related by the definition
\begin{equation}
  \vp \coloneqq \frac{\dot{\vq}}{Hf} \eqqcolon \vec{u}_{\mathrm{pec}}(\vq,\eta) \;.
\label{eq:peculiar_velocity_definition}
\end{equation}

In both KFT and SPT one then finds that the gravitational potential acting on particles or fluid elements with these momenta obeys Poisson's equation in the form \cite{peebles_large-scale_1980}
\begin{equation}
 \nabla_q^2 \tilde{\phi} = \frac{4 \pi G}{a^3 H^2 f^2} \left(\rho_\mm - \mpd_\mm\right) = \frac{3}{2 \mpd} \frac{\Omega_\mm}{f^2} \left(\rho - \mpd\right) \;,
\label{eq:poisson_equation}
\end{equation}
where $\rho_\mm = m \rho$ is defined \wrt to comoving volume elements, \ie the mean $\mpd_\mm = m \mpd$ is constant in time. Turning off interactions by setting this potential to zero thus ensures that the non-interacting regime in SPT and KFT is compatible.

In KFT, solving the linear Hamiltonian equation of motion for the phase-space coordinates $\vx = (\vq, \vp)$ gives the linear \emph{particle propagator}
\begin{equation}
 \mathcal{G}(\eta_1,\eta_2) = \matrix{cc}{
 \mathcal{I}_3 & \quad 2 \, \left(1-\e^{-\frac{1}{2}(\eta_1 - \eta_2)}\right) \, \mathcal{I}_3 \\
 0_3 & \e^{-\frac{1}{2}(\eta_1 - \eta_2)} \, \mathcal{I}_3
 } \heavi(\eta_1 - \eta_2)\;.
\label{eq:KFT_particle_propagator_cosmic_solution}
\end{equation}
We see that the expansion of the cosmological background drains away the initial momentum and thus particles asymptotically approach a finite comoving distance of $2\vpi$. Note that due to \eqref{eq:peculiar_velocity_definition} both $\vq$ and $\vp$ have the dimension of a length and therefore the free propagator is dimensionless.

In SPT we assume $\nabla \times \vec{u}_{\mathrm{pec}} = 0$ and use the Fourier transformed field doublet \eqref{eq:SPT_field_doublet} with $\theta \coloneqq \nabla_q \cdot \vec{u}_{\mathrm{pec}}$. The linear, non-interacting equation of motion for $\varphi_a \bigl(\vk,\eta\bigr)$ has the same structure \wrt to $\eta$ as that for the phase-space coordinates of KFT (see Appendix \ref{AppendixA}). Thus, the linear \emph{field propagator} is given by
\begin{equation}
\begin{split}
 g_{ab}(1,2) &= (2\pi)^3 \dirac\bigl(\vk_1 + \vk_2\bigr) \matrix{cc}{
 1 & \quad 2\left( 1 - \e^{-\frac{1}{2} (\eta_1 - \eta_2)} \right) \\
 0 & \e^{-\frac{1}{2} (\eta_1 - \eta_2)}
 } \heavi(\eta_1 - \eta_2) \\
 &\eqqcolon (2\pi)^3 \dirac\bigl(\vk_1 + \vk_2\bigr) g_{ab}(\eta_1, \eta_2) \;,
\end{split}
\label{eq:SPT_field_propagator_free_solution}
\end{equation}
where we use combined field arguments $1 = \bigl(\vk_1, \eta_1\bigr)$.

Both initial probability distributions, $\mathcal{P}(\txi)$ for KFT and $\mathcal{P}(\ini{\varphi_a})$ for SPT are fully determined once the auto- and cross-correlation functions of the density and velocity fields are specified. At sufficiently early times $\delta,\, u_{\mathrm{pec}} \ll 1$, and thus the linear limit of the continuity equation is an accurate description for the evolution of the density contrast. We insert the linear growing mode solution $\delta(\eta) = D_+(\eta) \, \ini{\delta}$ into the continuity equation to find
\begin{equation}
  D_+(\eta) \ini{\delta}(\vq) + \nabla_q \cdot \vec{u}_{\mathrm{pec}}(\vq,\eta) = 0 \;,
\label{eq:growing_mode_ini_condition1}
\end{equation}
where we have used $\md D_+ / \md \eta = D_+$. At initial time $\eta_i$ this then turns into
\begin{equation}
  \ini{\delta}(\vq) = -\nabla_q \cdot \vini{u}_{\mathrm{pec}}(\vq) = -\ini{\theta}(\vq) \;,
\label{eq:growing_mode_ini_condition2}
\end{equation}
where the normalisation $D_+(\eta_\mi) = 1$ was used. This equation relates all three necessary correlation functions to the initial density contrast power spectrum $\iniPS(k)$ (see Appendices \ref{AppendixA} and \ref{AppendixB}). This choice of initial fields puts the system into the linear growing mode of the gravitationally interacting regime.

The SPT perturbation series of any cumulant of the fields \eqref{eq:SPT_field_doublet} can be obtained as the sum of all connected Feynman diagrams built from 
\begin{equation}
\begin{split}
 \mytikz{  \draw[dline] (0,0) \nl{\varphi_a(1)} -- (\feyn,0) coordinate (v); \draw[pline] (v) -- (2\feyn,0) \nl{\chi_b(2)}; } &= \mi \, g_{ab}(1,2) \;, \qquad
 \mytikz{  \draw[psline] (0,0) \nl{\varphi_a(1)} -- (2\feyn,0) \nl{\varphi_b(2)};} = P^{\mlin}_{ab}(1,2) \;, \\
 \mytikz{ \draw[pline] (0,0) node[left] {$\chi_a(1)$} -- (\feyn,0) coordinate (v); \draw[dline] (v) -- ++(\feyn,\feyn) node[right] {$\varphi_b(2)$}; \draw[dline] (v) -- ++(\feyn,-\feyn) node[right] {$\varphi_c(3)$}; } &= \mi \, \gamma_{abc}(1,2,3) \;.
\end{split}
\label{eq:SPT_diagram_building_blocks}
\end{equation}
The details of this diagrammatic approach can be found in \cite{Matarrese, Crocce2006, Crocce2006a}. The linearly evolved power spectrum $P_{ab}^{\mlin}$ and the vertex tensor $\gamma_{abc}$ describing the (non-linear) free-streaming kinematics are given in Appendix  \ref{AppendixA}. Using the form of these diagrams one can explicitly show that in this naive perturbative scheme the order of $\iniPS$ to which one can obtain any cumulant is inextricably tied to the loop order in the free-streaming kinematics. However, we have shown in \cite{Fabis2018} that the exact free-streaming KFT cumulants are at any $n$-point level of arbitrary order in $\iniPS$. Therefore, writing down an exact (in the limits of the SPT approximations) free-streaming cumulant using SPT would require contributions from arbitrarily high loop orders.

\subsection{Recovering the 1-loop SPT density power spectrum from KFT} \label{sec04-02}

We will now show that expanding the exact expression for the free-streaming KFT density power spectrum to second order in $\iniPS$ gives the corresponding one-loop result of SPT. The calculation of exact free $n$-point phase-space density cumulants has been discussed at length in \cite{Fabis2018}. We therefore simply state the KFT result for the density-contrast two-point cumulant in Fourier space excluding shot noise contributions,
\begin{equation}
\begin{split}
  {}^{\textsc{kft}}\Gdd[0](1,2) =& \, \frac{{}^{\textsc{kft}}\Gf_{\rho\rho}(1,2)}{\bar{\rho}^2} \\
   =& \, (2\pi)^3 \dirac\bigl(\vk_1 + \vk_2\bigr) e^{-\frac{\sigma_p^2 \, k_1^2}{2} \, ( \mg_1^2 + \mg_2^2 )} \int_{\ini{q}_{12}} e^{-\mi\vk_1 \cdot \vqi_{12}} \Bigl[ \Bigl( 1 + C_{\delta_1 \delta_2} \\
  & - \mi (\mg_1 + \mg_2) \bigl(\vk_1 \cdot \vec{C}_{p_1 \delta_2}\bigr) - \mg_1 \, \mg_2 \, \bigl( \vk_1 \cdot \vec{C}_{p_1 \delta_2} \bigr)^2 \Bigr) e^{\mg_1 \, \mg_2 \, \vk^{\,\top}_1 C_{p_1 p_2} \vk_1} - 1 \Bigr] \,.
\end{split}
\label{eq:KFT_exact_free_powerspectrum}
\end{equation}
We introduced the short-hand notation $\mg_1 = \gqp(\eta_1, 0)$. The initial density correlation $C_{\delta_1 \delta_2}$ is the Fourier transform of the initial power spectrum $\iniPS$. The momentum correlation $C_{p_1 p_2}$ and the cross-correlation $\vec{C}_{\delta_1 p_2}$ can both be expressed in terms of $\iniPS$ once we adopt the relation \eqref{eq:growing_mode_ini_condition2}. In Appendix \ref{AppendixB} we use this to expand \eqref{eq:KFT_exact_free_powerspectrum} in $\iniPS$ up to second order. Note that the Gaussian damping term in front of the integral in \eqref{eq:KFT_exact_free_powerspectrum} also has to be expanded since $\sigma_p = C_{p_1p_1} = \mathcal{O}(\iniPS)$. The result in linear order is
\begin{equation}
 {}^{\textsc{kft}}\Gdd[0,1](1,2) = (2\pi)^3 \dirac\bigl(\vk_1 + \vk_2\bigr) \, (1 + \mg_1 ) \, (1 + \mg_2 ) \, \iniPS(k_1) \;.
\label{eq:KFT_linear_power_spectrum}
\end{equation}
The second-order contribution can be split into two parts as
\begin{equation}
 {}^{\textsc{kft}}\Gdd[0,2](1,2) = {}^{\textsc{kft}}\Gdd[0,2,\mSE](1,2) + {}^{\textsc{kft}}\Gdd[0,2,\mMC](1,2) \;,
\label{eq:KFT_second_order_ps_contribution}
\end{equation}
where the \emph{self-energy} contribution
\begin{equation}
 {}^{\textsc{kft}}\Gdd[0,2,\mSE](1,2) \coloneqq (2\pi)^3 \dirac\bigl(\vk_1 + \vk_2\bigr) \biggl( -\frac{\sigma_p^2 \, k_1^2}{2}  (\mg_1^2 + \mg_2^2) \biggr) \, (1 + \mg_1 ) \, (1 + \mg_2 ) \, \iniPS(k_1)
\label{eq:KFT_SE_term}
\end{equation}
results from the first-order term in the expansion of the Gaussian damping factor. The first term to involve actual \emph{mode-coupling} is given by
\begin{equation}
\begin{split}
 {}^{\textsc{kft}}\Gdd[0,2,\mMC](1,2) \coloneqq& \, (2\pi)^3 \dirac\bigl(\vk_1 + \vk_2\bigr) \int_h \iniPS(h) \iniPS(\Delta h) \, \mg_1 \mg_2 \\
 &\times \Bigl( 2F_{\mA}^2(\Delta\vh,\vh) + \, (\mg_1 + \mg_2) \, (F_{\mA}F_{\mB})(\Delta\vh,\vh) + \frac{\mg_1 \mg_2}{2} F_{\mB}^2(\Delta\vh,\vh) \Bigr) \;, 
\end{split}
\label{eq:KFT_mode_coupling_term}
\end{equation}
where $\Delta\vh = \vk_1 - \vh$ and we introduced the mode-coupling kernel functions
\begin{align}
 F_{\mA}\bigl(\vk_1,\vk_2\bigr) \coloneqq& \, \frac{1}{2} \, \Bigl(\alpha\bigl(\vk_1,\vk_2\bigr) + \alpha\bigl(\vk_2,\vk_1\bigr)\Bigr) =  1 + \frac{1}{2} \bigl(\vk_1 \cdot \vk_2 \bigr) \left( \frac{1}{k_1^2} + \frac{1}{k_2^2} \right)  \label{eq:FA_kernel_definition}\;, \\
 F_{\mB}\bigl(\vk_1,\vk_2\bigr) \coloneqq& \, F_{\mA}\bigl(\vk_1,\vk_2\bigr) + \beta\bigl(\vk_1,\vk_2\bigr) = \, 1 + \bigl( \vk_1 \cdot \vk_2 \bigr) \left( \frac{1}{k_1^2} + \frac{1}{k_2^2} \right) + \frac{\bigl(\vk_1 \cdot \vk_2 \bigr)^2}{k_1^2 k_2^2} \;.
 \label{eq:FB_kernel_definition}
\end{align}
The functions $\alpha$ and $\beta$ define the three-point vertex $\gamma_{abc}$ of SPT, see \eqref{eq:A1-10a}--\eqref{eq:A1-10c}. Both $F_{\mA}$ and $F_{\mB}$ are structurally very similar to the familiar SPT kernel $F_2$ \cite{Bernardeau2002} and we will in fact see in section \ref{sec04-04} that they combine into $F_2$ in the presence of gravity.

The evolved SPT density power spectrum is obtained by applying two functional derivatives \wrt $J_1$ to $\ln Z_{\mathrm{SPT}}$ in \eqref{eq:A1-13}. The linear contribution is given by the tree-level of the perturbation series, \ie the $a=b=\delta$ component of \eqref{eq:A1-14}. After inserting \eqref{eq:SPT_field_propagator_free_solution} we find agreement between KFT and SPT at the linear level,
\begin{equation}
 {}^{\textsc{spt}}\Gdd[0,\mlin](1,2) = {}^{\textsc{kft}}\Gdd[0,1](1,2) \;.
\label{eq:SPT_free_linear_power_spectrum}
\end{equation}
The second-order terms in SPT can be obtained by computing all possible one-loop diagrams with two external $\varphi_1$-legs. Actually calculating them is straightforward but tedious and we therefore refer the reader to the usual SPT literature, \eg \cite{Crocce2006a}.
There are in total three one-loop diagrams. One of them represents the effects of mode-coupling,
\begin{equation}
  {}^{\textsc{spt}}\Gdd[\mMC](1,2)
   \coloneqq 2 \mytikz{ \draw[dline] (0,0) \nl{\varphi_\delta(1)} -- ++ (\feyn,0) coordinate (c1); \draw[pline] (c1) -- ++ (\feyn,0) coordinate (vl) ; \draw[psline] (vl) arc(180:0:\feyn) coordinate (vr) ; \draw[pline] (vr) -- ++ (\feyn,0) coordinate (c2); \draw[dline] (c2) -- ++ (\feyn,0) \nl{\varphi_\delta(2)} ; \draw[psline] (vr) arc(0:-180:\feyn) ; } \;.
\label{eq:SPT_mode_coupling_diagram}
\end{equation}
In the case of free-streaming we again need to plug in the corresponding propagator \eqref{eq:SPT_field_propagator_free_solution} and find agreement with KFT,
\begin{equation}
 {}^{\textsc{spt}}\Gdd[0,\mMC](1,2) = {}^{\textsc{kft}}\Gdd[0,2,\mMC](1,2) \;.
\label{eq:SPT_free_mode_coupling_term}
\end{equation}
The other two of the diagrams only differ by exchange of the external arguments and amount to
\begin{equation}
 {}^{\textsc{spt}}\Gdd[\mSE](1,2)
  \coloneqq 4 \mytikz{ \draw[dline] (0,0) \nl{\varphi_\delta(1)} -- ++ (\feyn,0) coordinate (c1); \draw[pline] (c1) -- ++ (\feyn,0) coordinate (vl) ; \draw[psline] (vl) arc(180:0:\feyn) coordinate (vr) ; \draw[psline] (vr) -- ++ (2\feyn,0) \nl{\varphi_\delta(2)} ; \draw[pline] (vr) arc(0:-90:\feyn) coordinate (c2); \draw[dline] (c2) arc(-90:-180:\feyn) ; } 
 + 4 \mytikz{ \draw[psline] (0,0) \nl{\varphi_\delta(1)} -- ++ (2\feyn,0) coordinate (vl) ; \draw[psline] (vl) arc(180:0:\feyn) coordinate (vr) ; \draw[pline] (vr) -- ++ (\feyn,0) coordinate (c1); \draw[dline] (c1) -- ++ (\feyn,0) \nl{\varphi_\delta(2)} ; \draw[dline] (vr) arc(0:-90:\feyn) coordinate (cl); \draw[pline] (cl) arc(-90:-180:\feyn) ; } \;.
\label{eq:SPT_self_energy_diagram}
\end{equation}
Again we plug in the free-streaming propagator \eqref{eq:SPT_field_propagator_free_solution} and find the correspondence
\begin{equation}
 {}^{\textsc{spt}}\Gdd[0,\mSE](1,2) = {}^{\textsc{kft}}\Gdd[0,2,\mSE](1,2) \;.
\label{eq:SPT_free_self_energy_term}
\end{equation}

These results reinforce the notion that in the absence of gravity, KFT automatically resums all perturbative corrections due to non-linear kinematics in SPT. For the damping factor in \eqref{eq:KFT_exact_free_powerspectrum} we can calculate this directly: The self-energy diagrams in \eqref{eq:SPT_self_energy_diagram} and their equivalents in higher loop-orders can be understood as corrections to the linear SPT propagator \eqref{eq:SPT_field_propagator_free_solution} that can be resummed into a damping factor. Applying the usual SPT propagator resummation in the large-$k$ regime \cite{Matarrese,Crocce2006a,Bernardeau2012} to \eqref{eq:SPT_field_propagator_free_solution} yields\footnote{This resummation can be applied here in the non-interacting case since it only relies on the form of the vertices $\gamma_{abc}$ and the propagator property $g_{ab}(\eta_1,\eta_2) \, g_{bc}(\eta_2,\eta_3) = g_{ac}(\eta_1,\eta_3) \, \theta(\eta_1-\eta_2) \, \theta(\eta_2-\eta_3)$, both of which are unchanged.}
\begin{equation}
 g_{ab}^{\mathrm{res}}(1,2) = g_{ab}(1,2) \, \exp\biggl\{-\frac{\sigma_p^2 \, k_1^2}{2} (\mg_1 - \mg_2)^2\biggr\} \;.
\label{eq:SPT_resummed_free_propagator}
\end{equation}
Replacing the bare or `tree-level' linear propagator $g_{ab}$ in the linear SPT power spectrum \eqref{eq:A1-14} with the resummed $g_{ab}^{\mathrm{res}}$ then leads to complete agreement with the fully damped KFT result excluding any mode-coupling contributions,
\begin{equation}
\begin{split}
 {}^{\textsc{spt}}\Gdd[0,\text{res}](1,2) &= {}^{\textsc{kft}}\Gdd[0,\text{no MC}](1,2) \\
 &= (2\pi)^3 \dirac\bigl(\vk_1 + \vk_2\bigr) \, (1 + \mg_1 ) \, (1 + \mg_2 ) \, e^{-\frac{\sigma_p^2 \, k_1^2}{2} \, ( \mg_1^2 + \mg_2^2 )}\iniPS(k_1) \;.
\end{split}
\label{eq:KFT_damped_linear_power_spectrum}
\end{equation}
Fully damped KFT results including some degree of mode-coupling cannot be reproduced in this simple manner: If we, for instance, replace all linear propagators in the defining diagram of the SPT 1-loop MC contribution \eqref{eq:SPT_mode_coupling_diagram} with $g_{ab}^{\mathrm{res}}$, it will lead to integrals over its damping factor.

One should also note that while similar, the physical reasoning behind the damping factor is slightly different between the two theories. When considering the large-$k$ regime of small wavelengths in SPT, one can approximate their coupling to long wavelength modes by treating the latter as a random background \cite{Bernardeau2012}. In this so-called \emph{eikonal} approximation the resummed propagator takes the form
\begin{equation}
 g_{ab}^{\mathrm{res}}(1,2) = g_{ab}(1,2) \, \biggl\langle \exp\biggl\{\int_{\eta_2}^{\eta_1} \md \eta \int_{h \ll k_1} \frac{\vk_1 \cdot \vh}{h^2} \, \theta(\vh,\eta)\biggr\} \biggr\rangle_{\mF} \;,
 \label{eq:SPT_resummed_propagator_eikonal_approximation}
\end{equation}
where the argument of the exponential is the projection of the time-integrated velocity field along $\vk_1$, and the Fourier integral runs only over the large-scale modes $h \ll k_1$. If the large-scale velocity is assumed to evolve linearly, $\theta(\vh, \eta) = g_{2a}(\eta,0) \, \ini{\phi_a}(\vh)$, and we employ the usual Gaussian initial conditions, then the field-averaged exponential in \eqref{eq:SPT_resummed_propagator_eikonal_approximation} is reduced to the Gaussian damping factor in \eqref{eq:SPT_resummed_free_propagator} involving only the local velocity variance. By introducing this separation of scales, one however has to formally define the velocity variance as
\begin{equation}
 \sigma_p^2(k_1) = \frac{1}{3} \int_{h \ll k_1} \frac{\iniPS(h)}{h^2} \;.
\label{eq:SPT_velocity_variance_integral_with_cutoff}
\end{equation}

In KFT the dynamical origin of the damping factor is not a large-scale background of random flows, but rather the fact that the Gaussian initial momentum field is sampled by an ensemble of $N$-particle systems, thereby imprinting the local one-point variance of this field onto single particles as momentum dispersion between different ensemble members. In the ensemble average this leads to an uncorrelated component of random motion that damps existing correlations \cite{Dombrowski2018}. In contrast to SPT, no separation of scales is necessary, so the damping factor is valid for any mode $k_1$ and the integral in \eqref{eq:SPT_velocity_variance_integral_with_cutoff} runs over all modes.

\subsection{Density bispectrum} \label{sec04-04}
The contribution to the free-streaming density-contrast bispectrum\footnote{excluding shot-noise contributions} in KFT is given by three-particle correlations. We present the important steps of its calculation in Appendix \ref{App3}, while a detailed discussion can be found in \cite{Fabis2018}. The final result up to second order in $\iniPS$ is given by
\begin{align}
 {}^{\textsc{kft}}\Gf[2]_{\delta \delta \delta}(1,2,3) &=  (2\pi)^3 \dirac\bigl(\vk_1 + \vk_2 + \vk_3\bigr) e^{-\frac{\sigma_p^2}{2} \left( \mg_1^2 \, k_1^2 + \mg_2^2 \, k_2^2 + \mg_3^2 \, k_3^2\right)} \biggl[ \iniPS(k_1) \, \iniPS(k_2) ( 1 + \mg_1 )\nonumber \\
 &\qquad \times ( 1 + \mg_2 ) \, \Bigl(  \mg_3 \, 2 F_{\mA}\bigl(\vk_1,\vk_2\bigr) + \mg_3^2 \, F_{\mB}\bigl(\vk_1,\vk_2\bigr) \Bigr) + \mathrm{cycl.~perm.} \biggr] \,.
\label{eq:KFT_free_quadratic_bispectrum}
\end{align}
At the initial time $\eta_1 = \eta_2 = \eta_3 = 0$, the bispectrum vanishes since the initial density field is Gaussian. At any later time however, the bispectrum will build up power. This shows that not only gravity, but any non-linear kinematic effect will force an initially Gaussian density field with momentum correlations to develop non-Gaussian features.

In SPT this effect is apparent since the lowest-order tree-level contribution to the bispectrum must necessarily be constructed from the $\gamma_{abc}$-vertex and reads
\begin{equation}
\begin{split}
 {}^{\textsc{spt}}\Gf[\mathrm{tree}]_{\delta \delta \delta}(1,2,3) =& \, 2 \mytikz{ \draw[dline] (0,0) \nl{\varphi_\delta(3)} -- ++ (\feyn,0) coordinate (c1); \draw[pline] (c1) -- ++ (\feyn,0) coordinate (v); \draw[psline] (v) -- ++ (2\feyn,2\feyn) node[right] {$\varphi_\delta(1)$} ; \draw[psline] (v) -- ++ (2\feyn,-2\feyn) node[right] {$\varphi_\delta(2)$} ;} + \mathrm{cycl.} \\
 =& \, (2\pi)^3 \dirac\bigl(\vk_1 + \vk_2 + \vk_3\bigr) \biggl[  \iniPS(k_1) \iniPS(k_2)(1 + \mg_1) (1 + \mg_2) \\
 &\times \Bigl(  \mg_3 \, 2 F_{\mA}\bigl(\vk_1,\vk_2\bigr) + \mg_3^2 \, F_{\mB}\bigl(\vk_1,\vk_2\bigr) \Bigr) + \mathrm{cycl.~perm.} \biggr] \;.
\end{split}
\label{eq:SPT_free_tree_bispectrum}
\end{equation}
Not surprisingly, this corresponds to the KFT result \eqref{eq:KFT_free_quadratic_bispectrum} without the damping factor, further reinforcing the notion that the corresponding KFT terms represent a resummation of SPT terms in the absence of gravity. As for the mode-coupling contribution to the power spectrum, damping effects cannot appear here because the bispectrum is itself already of second order in $\iniPS$. 

\subsection{Effects of gravity} \label{sec04-05}
It is apparent that the above results for the density spectra, in terms of their Fourier kernel structure, look very similar to their well-known SPT counterparts in the presence of gravity. The reason for this is the assumption of a vorticity-free velocity field, leading to the Fourier space relation
\begin{equation}
 \vec{u}_{\mathrm{pec}}(\vk) = \frac{\mi \vk}{k^2} \theta(\vk) \;.
\label{eq:vorticity_free_velocity_field_fourier_space}
\end{equation}
In KFT this relation is only used at the initial time to express correlations in terms of Fourier integrals of $\iniPS$ alone. These integrals are subsequently coupled by the exact solution of the free-streaming kinematics without any restriction on vorticity.

In SPT, however, \eqref{eq:vorticity_free_velocity_field_fourier_space} enters the kinematic evolution itself because it is used to define the vertex in \eqref{eq:SPT_diagram_building_blocks}. Since this evolution is solved perturbatively, the vertices prescribe an integration over time, which due to the knowledge of the exact kinematic solution does not appear in KFT. In the absence of gravity these integrals combine with the propagator components to reproduce the KFT results. However, when including gravity in SPT we have to change the linear propagator into its well-known form \cite{Bernardeau2002}
\begin{equation}
 g_{ab}(\eta_1,\eta_2) = \left( \frac{\e^{\eta_1-\eta_2}}{5} \matrix{cc}{ 3 & 2 \\ 3 & 2 } + \frac{\e^{-\frac{3}{2}(\eta_1-\eta_2)}}{5} \matrix{cc}{ 2 & -2 \\ -3 & 3 } \right) \heavi(\eta_1 - \eta_2) \;.
\label{eq:SPT_field_propagator_interacting_solution}
\end{equation}
There are two important features to note in comparison to \eqref{eq:SPT_field_propagator_free_solution}. First, the $g_{21}$ component no longer vanishes since it describes how overdensities $\varphi_\delta$ accelerate their surroundings by means of their gravitational pull, thus acting as sources for $\varphi_\theta = -\nabla \cdot \vec{u}_{\mathrm{pec}}$. And second, we see that both fields $\varphi_a$ now have the same scaling with time, albeit with different prefactors. For the linear power spectrum matrix they combine into
\begin{equation}
  {}^{\textsc{spt}}G^{\mlin}_{ab}(1,2) = (2\pi)^3 \, \dirac\bigl(\vk_1 + \vk_2\bigr) \, e^{\eta_1 + \eta_2} \, \iniPS(k_1) \matrix{cc}{ 1 & 1 \\ 1 & 1}\;.
\label{eq:SPT_interacting_linear_power_spectrum}
\end{equation}

In the case of the mode-coupling term and the bispectrum, $g_{21}$ only enters through \eqref{eq:SPT_interacting_linear_power_spectrum}. Thus when written in terms of $g_{ab}$, $P_{ab}^{(\mlin)}$ and $\gamma_{abc}$, they still have the same form as in the non-interacting regime. In both cases the integrals over time lead to a global growth behaviour in time, but with residual constant prefactors weighting the kernels. For the mode-coupling term \eqref{eq:SPT_mode_coupling_diagram} the overall time-scaling will be $\e^{2\eta_1 + 2\eta_2}$ and it turns out that the kernel can again be written in terms of $F_\mA$ and $F_\mB$ as \cite{Jain1994}
\begin{equation}
 2 \left( \frac{5}{7}  F_{\mA} + \frac{2}{7} \left( F_{\mB} - F_{\mA} \right) \right)^2 = 2 \left( \frac{3}{7} F_{\mA} + \frac{2}{7} F_{\mB} \right)^2 = 2 \bigl(F_2^{(\mathrm{s})}\bigr)^2 \;.
\label{eq:SPT_interacting_mode_coupling_kernel}
\end{equation}
Here, $F_2^{(\mathrm{s})}$ denotes the familiar symmetrised second-order perturbation kernel \cite{Bernardeau2002}, and all kernels are evaluated at $\vk_1 \rightarrow \Delta\vh = \vk_1 - \vh$ and $\vk_2 \rightarrow \vh$. For the bispectrum the overall time scaling of the diagram shown in \eqref{eq:SPT_free_tree_bispectrum} is $\e^{\eta_1 + \eta_2 + 2\eta_3}$ and the kernel is
\begin{equation}
 2 \left( \frac{5}{7} F_{\mA} + \frac{2}{7} \left( F_{\mB} - F_{\mA} \right) \right) = 2 \left( \frac{3}{7} F_{\mA} + \frac{2}{7} F_{\mB} \right) = 2 F_2^{(\mathrm{s})} \;.
\label{eq:SPT_interacting_bispectrum_kernel}
\end{equation}
This shows that the peculiar constant prefactors appearing in SPT are rooted in the linear growth behaviour \eqref{eq:SPT_field_propagator_interacting_solution} of individual field modes and not in the non-linear kinematics that lead to mode-coupling. In the case of the self-energy term \eqref{eq:SPT_self_energy_diagram}, the $g_{21}$-component will change the kernel to the symmetrised SPT kernel $F_3^{(\mathrm{s})}$ \cite{Makino1992,Jain1994,Bernardeau2002}, which precludes any simple form like \eqref{eq:SPT_free_self_energy_term}. One can however consider its limit in the large-$k$ regime which turns out to be
\begin{equation}
 {}^{\textsc{spt}}G^{(\mathrm{SE})}_{\delta \delta}(1, 2) \underset{k_1 \gg h}{\approx} {}^{\textsc{spt}}G^{(\mlin)}_{\delta \delta}(1, 2) \biggl(- \frac{k_1^2 \sigma_p^2}{2} \, \left(\e^{2\eta_1} + \e^{2\eta_2} \right) \biggr) \;.
\label{eq:SPT_interacting_self_energy_term}
\end{equation}
This is not surprising since we can also perform the large-$k$ resummation of the SPT propagator in the interacting case \cite{Matarrese,Crocce2006a,Bernardeau2012}, turning the linear power spectrum into
\begin{equation}
 {}^{\textsc{spt}}G_{\delta \delta}^{(\mathrm{lin,res})}(1, 2) = (2\pi)^3 \, \dirac\bigl(\vk_1 + \vk_2\bigr) \,  e^{-\frac{\sigma_p^2 k_1^2}{2}  (e^{2\eta_1} + e^{2\eta_2} )} \, e^{\eta_1 + \eta_2} \, \iniPS(k_1)  \;.
\label{eq:SPT_interacting_resummed_linear_power_spectrum}
\end{equation}
A first-order expansion of the damping factor then leads back to \eqref{eq:SPT_interacting_self_energy_term}.

We can almost recover \eqref{eq:SPT_interacting_resummed_linear_power_spectrum} from the KFT expression \eqref{eq:KFT_damped_linear_power_spectrum} if we simply force particles onto Zel'dovich trajectories by replacing $\gqp(\eta_1, \eta_2) \rightarrow D_+(\eta_1) - D_+(\eta_2)$, such that $\mg_1 \rightarrow e^{\eta_1} -1$. As we argued in \cite{Bartelmann2016}, free KFT with Zel'dovich trajectories is essentially equivalent to first-order Lagrangian Perturbation Theory (LPT). It is thus not surprising that we reproduce the damped growth behaviour of the linear power spectrum, since it is well known that LPT and large-$k$ resummed SPT coincide in the linear regime \cite{Matsubara2008}.

The same is however not true for the mode-coupling term and the bispectrum. The KFT results \eqref{eq:KFT_mode_coupling_term}, \eqref{eq:KFT_free_quadratic_bispectrum} are composed of terms with differing growth behaviours, because they originate from a mix of initial correlations between density and momentum field, which scale differently in time when extrapolated with the inertial particle trajectories. As seen in \eqref{eq:SPT_field_propagator_interacting_solution}, \eqref{eq:SPT_interacting_linear_power_spectrum}, the inclusion of gravity in SPT leads to the same linear growth behaviour for all three kinds of initial correlations, which results in the global growth behaviour of the corresponding SPT terms. One can obtain the SPT power spectrum up to 1-loop order with an overall damping factor the same as in \eqref{eq:SPT_interacting_resummed_linear_power_spectrum}, by considering third-order LPT expanded to second order in the linear power spectrum $D_+^2 \, \iniPS$ \cite{Matsubara2008}. This suggests that in order to achieve comparable results in KFT one has to go into perturbation theory to include effects of particles being diverted from their inertial trajectories due to gravitational interaction, since the order of LPT coincides with the polynomial order of curves particles can move on.

We want to stress again that the advantage of KFT exploited in \cite{Bartelmann2017,Dombrowski2018} is that in the absence of gravity it is a resummation of the non-linear kinematics of SPT and one can deduce exact cumulant expressions like \eqref{eq:KFT_exact_free_powerspectrum}. The point is to calculate these \emph{without} expanding in orders of $\iniPS$, which would correspond to the SPT notion of performing perturbation theory in orders of the linear solution $\delta^{(\mlin)}(t)$ of the density contrast. For times with $\delta^{(\mlin)}(t) > 1$, this expansion will break down. Combining this resummation of free kinematics with appropriate Zel'dovich-type trajectories \cite{Bartelmann2015a}, which greatly reduce the amplitude of the potential acting relative to them on large scales and for late cosmic times, should therefore perform much better than SPT at low perturbative orders. This was confirmed by the results in \cite{Bartelmann2016}. Due to the similar physical picture behind the perturbative expansion it could prove worthwhile to investigate how this scheme corresponds to Lagrangian perturbation schemes in future work.

\begin{figure}
    \centering
    \includegraphics{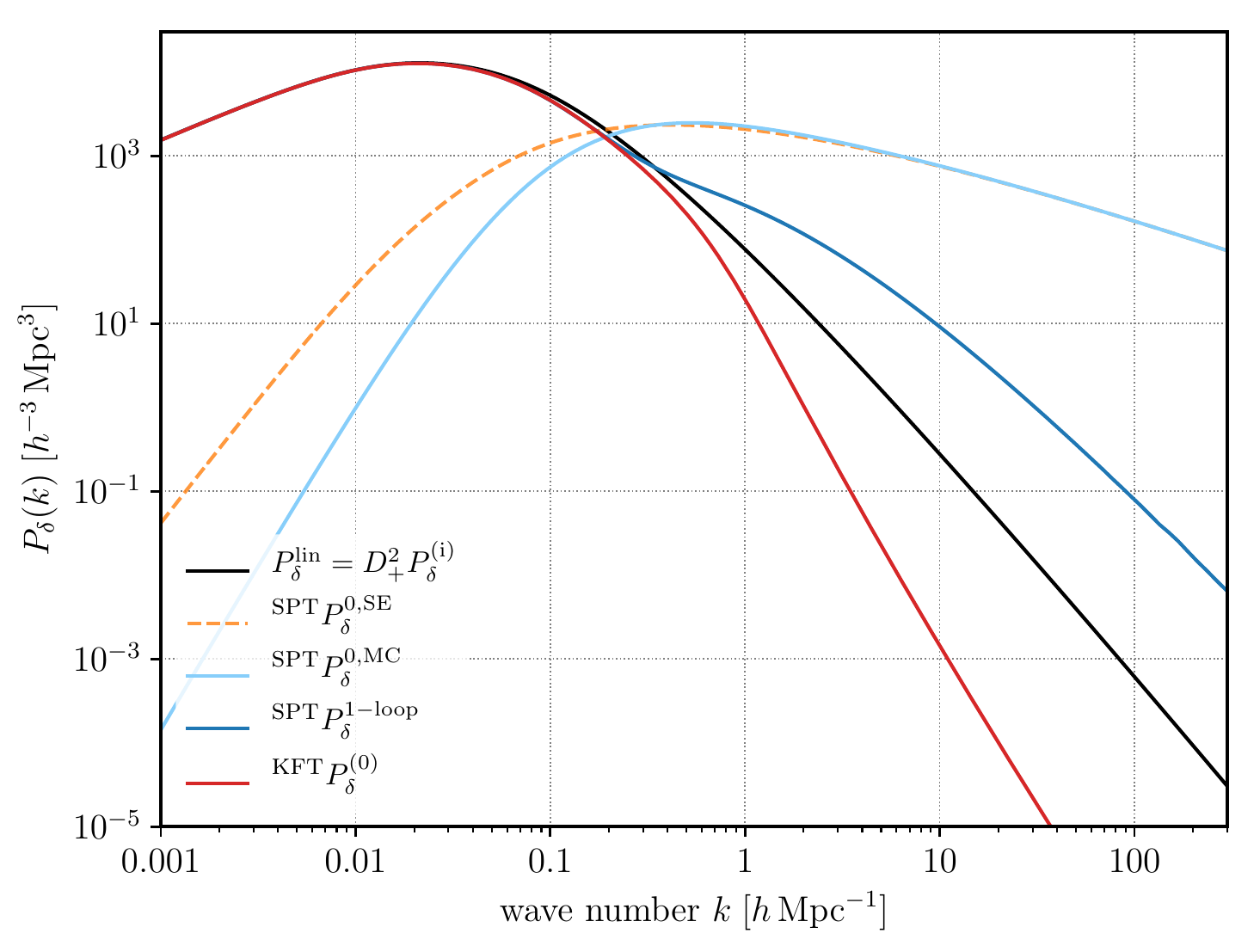}
    \caption{Comparison of free-streaming power spectra in SPT and KFT, both for a Zel'dovich propagator value corresponding to $a=1$. Shown are the 1-loop SPT power spectrum (dark blue) computed from the sum of \eqref{eq:SPT_free_linear_power_spectrum} (black), \eqref{eq:SPT_mode_coupling_diagram} (light blue) and \eqref{eq:SPT_self_energy_diagram} (orange), as well as the full free-streaming KFT spectrum computed from  \eqref{eq:KFT_exact_free_powerspectrum} (red). Dashed lines indicate negative values.}
    \label{fig:all_spectra}
\end{figure}

\subsection{Illustrating the effects}
Finally, we will illustrate the differences between the KFT and SPT power spectra in the free-streaming regime. Due to the absence of gravitational interactions, the only effects that lead to a deviation of the power spectrum from the linear evolution arise from the initial momentum correlations. For purely Newtonian dynamics with the propagator given by \eqref{eq:KFT_particle_propagator_cosmic_solution}, these effects would appear – from a cosmological perspective – on extremely small scales of the order of $k > 10^{5} \, h\,\mathrm{Mpc}^{-1}$ because the propagator can maximally reach a value of $g_{qp} \approx 2$ due to the expanding background. However, as explained in \autoref{sec04-05}, replacing the bounded Newtonian propagator by the unbounded Zel'dovich propagator $\gqp(\eta_1, \eta_2) = D_+(\eta_1) - D_+(\eta_2)$ allows us to capture part of the gravitational interactions in a way that reproduces the linear growth of the density contrast on large scales. Using this propagator, the effects due to initial correlations including the damping appear well within the scale-regime relevant for cosmology, as can be seen for the power spectra
\begin{equation}
 P_\delta(k_1, \eta_1) \coloneqq \int_{k_2} \Gdd(1,2) \, \Bigr|_{\eta_2 = \eta_1}
\end{equation}
shown in Fig.\ \ref{fig:all_spectra}.

For the free-streaming 1-loop SPT result, we see the usual cancellation between the leading-order contributions of ${}^{\textsc{spt}}\Gf[\mathrm{MC}]_{\delta \delta}$ and ${}^{\textsc{spt}}\Gf[\mathrm{SE}]_{\delta \delta}$ at small scales also found in interacting SPT \cite{Matsubara2008,Vishniac1983}. Considering that the self-energy term in SPT is the first-order approximation of the full damping factor in KFT, this suggests that mode-coupling terms counter the effects of damping. In fact it was shown in \cite{Blas2013} that by taking the appropriate mode-coupling terms into account when resumming the SPT power spectrum in the large-$k$ regime, the exponential damping factor of the propagator resummation is exactly cancelled. Since this holds true for any type of propagator, it applies to the non-interacting regime as well. 

\begin{figure}
    \centering
    \includegraphics{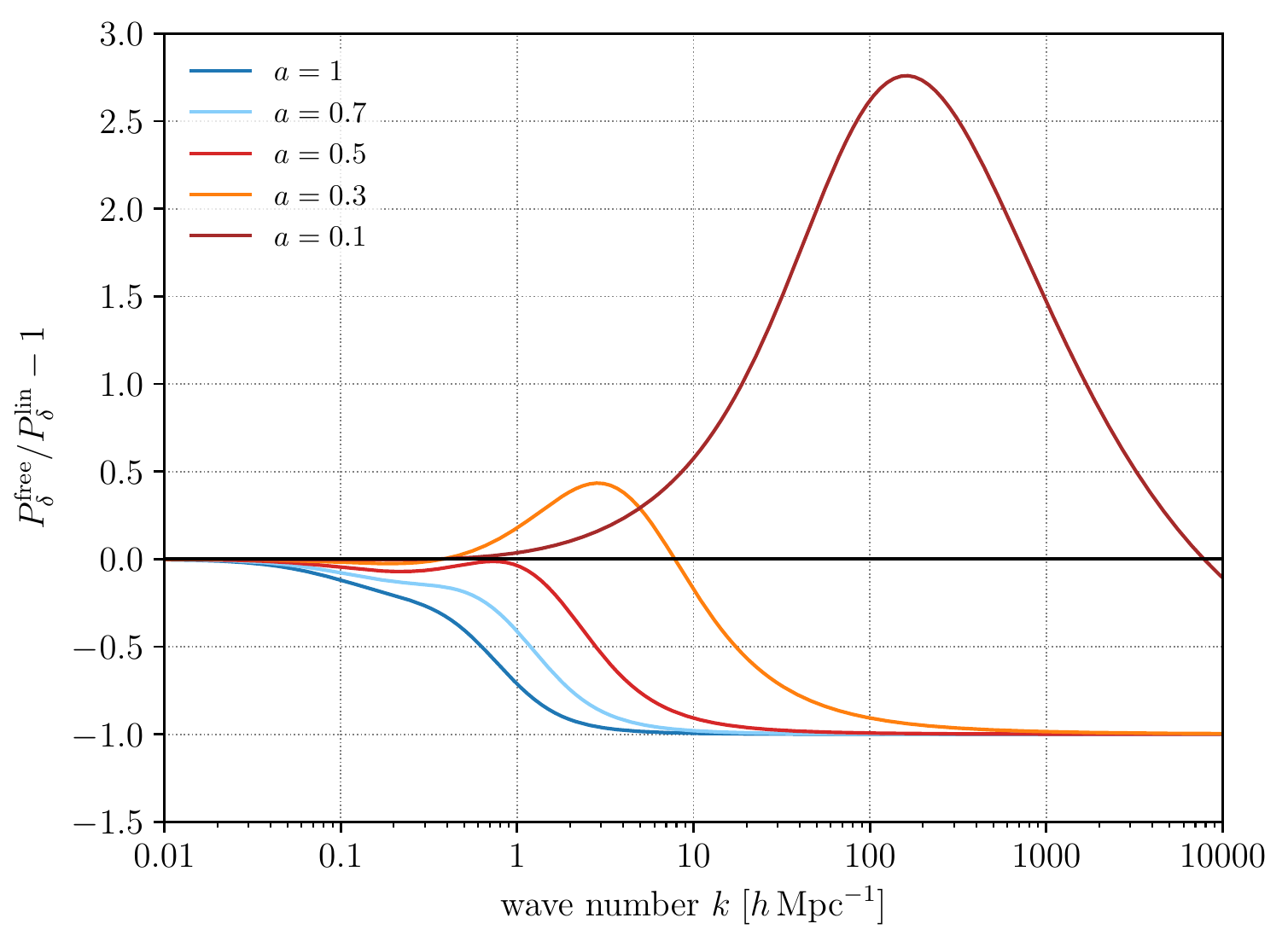}
    \caption{Relative deviation of the free-streaming KFT spectrum computed from \eqref{eq:KFT_exact_free_powerspectrum} to the linear power spectrum for Zel'dovich propagator values corresponding to different scale factors. The spectra display transient effects.}
    \label{fig:transient}
\end{figure}

In the full free-streaming KFT power spectrum obtained from \eqref{eq:KFT_exact_free_powerspectrum}, shown in Fig.\ \ref{fig:all_spectra}, we observe a similar cancellation of the exponential damping factor, leaving only a residual suppression of structures on small scales. In \cite{Bartelmann2017,Dombrowski2018} we have shown that one can understand this as a balancing process between diffusive local momentum variance and non-local momentum correlations that lead to convergent flows due to the continuity relation \eqref{eq:growing_mode_ini_condition2}. For the exact free-streaming power spectrum this leads to transient behaviours shown in Fig.\ \ref{fig:transient}, where depending on the elapsed time and the scales considered, the power of density fluctuations is either enhanced or damped. This suggests that KFT not only resums free-streaming SPT, but includes additional effects that are not captured by SPT due to its approximated dynamical equations. We conjecture that this is a first example where the effects of crossing streams can be observed, because without gravity binding particles in regions of convergent flows they will simply fly by each other and disperse again.

When gravity is switched on,\footnote{in addition to the gravitational effects already contained in the Zel'dovich trajectories.} it therefore needs to balance the damping effect and build up additional structures. If we expand the interaction operator $\e^{\hat{\Phi} \cdot \sigma \cdot \hat{\Phi}}$ in the KFT generating functional \eqref{eq:KFT_GF_operator_version} only to first order in the interaction potential, the damping term still dominates the non-linear power spectrum on small scales, and not enough structure can be built up. We have found in \cite{Bartelmann2016} that in order to achieve results close to the non-linear power spectrum known from simulations, we need to expand the damping term as well. However, we can exploit the knowledge of the full damping of the theory to avoid perturbative approximations in the interaction potential and instead find non-perturbative expressions such as the one proposed in \cite{bartelmann2020}. Alternatively, the exponential form of the damping terms can serve as a guide for resummation schemes.

If one does not wish to mix effects of gravity into the free theory, as the Zel'dovich trajectories implicitly do, one has to perturb the inertial trajectories described by \eqref{eq:KFT_particle_propagator_cosmic_solution} with the potential \eqref{eq:poisson_equation}. But a finite order of perturbative terms will obviously not lead to the linear growth of SPT, where gravity is a source in the linear equations of motion whose solution must thus be considered to be of infinite order in the interaction potential. However, we have shown in \cite{Lilow_2019} that it is possible to rewrite the KFT generating functional with the macroscopic phase-space density as the main dynamical object, while still preserving the full underlying microscopic dynamics. This automatically resums the subclass of terms in the perturbative expansion of $\e^{\hat{\Phi} \cdot \sigma \cdot \hat{\Phi}}$ representing self-similar growth and reproduces the large-scale linear growth of SPT at tree-level.

\section{Summary and conclusions} \label{sec05}

In this paper we made a first comparison between Kinetic Field Theory \cite{Bartelmann2016,Bartelmann2017,Viermann2015,Dombrowski2018} and Eulerian Standard Perturbation Theory as models for cosmic large-scale structure formation. Both approaches allow us to follow the time evolution of $n$-point cumulants of macroscopic fields by encapsulating both, their statistics and dynamics into a generating functional. However, we have shown that the underlying microscopic dynamics are treated very differently by the two approaches. Starting with the exact description of the system in terms of $N$-particle phase-space dynamics we could then show the following:

\begin{itemize}
 \item Both theories use a coarse-graining ensemble average to establish the connection with the macroscopic fields. However, in KFT we use it at the initial time only and then follow the actual Hamiltonian phase-space dynamics of the particle ensemble. In SPT, on the other hand, the coarse-graining is applied to the entire dynamical evolution. A truncation of the resulting BBGKY-hierarchy at first order leads to the Vlasov-Poisson equations and the treatment of the cosmic matter distribution as a collisionless system. This ab initio neglect of small-scale dynamics is absent in KFT. But since we take the thermodynamic limit $N \rightarrow \infty$ in cosmology, these subdominant shot-noise contributions can be neglected in the final results for $n$-point cumulants in KFT. In this limit KFT is equivalent to the Vlasov-Poisson description.

 \item Integrating out momentum information from the Vlasov equation leads to a hierarchy of fluid equations. KFT encompasses this entire hierarchy by construction. In SPT, however, one introduces the macroscopic velocity field and its dispersion tensor, where the latter is set to zero throughout the evolution, thus truncating the hierarchy. This imposes the existence of a single-valued velocity field at any point and time. Due to this single-stream approximation, SPT cannot describe crossing streams in the matter field, which necessarily occur when non-linear structures start to form.

 \item Furthermore, the hierarchy of fluid equations obtained from the Vlasov equation suggests perturbation theory in orders of the interaction potential. In SPT the velocity field is employed as a dynamical variable, which switches the roles of free-streaming kinematics and gravitational interaction as linear and non-linear terms in the equations of motion (see also \cite{Viermann2015}). Perturbation theory in SPT is thus an expansion in the kinematics. KFT, on the other hand, expands in the interactions, or more precisely, the deviations of particles from inertial trajectories due to these interactions. 
\end{itemize}

This difference between perturbative schemes suggested the non-interacting regime as the best starting point for a quantitative comparison of $n$-point cumulants of the cosmological density field. We have shown in \cite{Fabis2018} that in this regime KFT produces exact expressions for any such cumulant, since the free evolution of the $N$-particle ensemble is easily solved. Once we expanded the matter power spectrum and bispectrum in orders of the initial density contrast power spectrum $P^{(i)}_\delta(k)$, we found agreement with the corresponding SPT results. This suggests that in the absence of gravity the KFT result contains a resummation of all terms appearing in SPT. However, since KFT is free of the SSA and the assumption of vanishing vorticity in the velocity field, it includes effects beyond SPT even in the free regime as shown in Fig.\ \ref{fig:transient}. The appearance of such transient behaviours in the matter power spectrum, also found in \cite{Bartelmann2017,Dombrowski2018}, can be interpreted as a first sign of crossing streams.

We furthermore discussed how the inclusion of gravity changes the expressions for the cumulants. We saw that the similarities in the Fourier kernel structure between the two theories is mostly untouched since it is rooted in the assumption of vanishing vorticity in the (initial) velocity field. What changes are mainly the relative constant prefactors in these kernels introduced by the linear propagator of interacting SPT. While the linear large-scale growth of the power spectrum can be easily reproduced in KFT by forcing particles onto Zel'dovich trajectories, direct comparisons with non-linear corrections of SPT will require the inclusion of KFT perturbative corrections due to gravitational interactions.

So far, two different approaches to perturbation theory have been explored in KFT. The simplest one is an expansion in the interaction operator in \eqref{eq:KFT_GF_operator_version} in orders of the interaction matrix $\sigma$ \cite{Bartelmann2016}. Another approach is the reformulation of the theory in terms of the phase-space density as the main dynamical object \cite{Lilow_2019}. This approach is structurally much more similar to SPT and should thus be better suited when it comes to understanding the relationship between KFT and SPT in the interacting case. 

Finally, we pointed out that the fact that the free theory of KFT includes the full knowledge of the non-interacting system -- and in particular the form of the full damping factor -- can be used to successfully construct non-perturbative solutions for the full, interacting system.

\acknowledgments{
We are grateful for many helpful comments and discussions to Carsten Littek, Celia Viermann, Johannes Dombrowski, Sara Konrad and Bj\"orn Malte Sch\"afer. This work was supported in part by Deutsche Forschungsgemeinschaft (DFG) under the German Excellence Initiative, the Collaborative Research Centre TR 33 ``The Dark Universe'' and the German Excellence Strategy EXC-2181/1 - 390900948 (the Heidelberg STRUCTURES Excellence Cluster), and by a Technion fellowship.}

\appendix

\section{CDM dynamics in KFT and SPT} \label{AppendixA}

We first consider the phase-space dynamics of cold dark matter particles in KFT. On an expanding FLRW
background, they are described by an appropriate single-particle Lagrange function and peculiar gravitational potential \cite{Bartelmann2015a}
\begin{equation}
 L(\vq,\dot{\vq},t) = \frac{m}{2} a^2 \dot{\vq}^{\,2} - m \phi(\vq,t) \;, \qquad \nabla_q^2 \phi = \frac{4 \pi G}{a}  \left(\rho_\mm - \mpd_\mm\right) \;,
\label{eq:KFT_Lagrangian}
\end{equation}
with \textit{comoving} coordinates $\vq = \vec{r} / a$ and \textit{comoving} matter density $\rho_\mm$, \ie the mean density $\mpd_\mm$ is constant in time. Repeating the steps detailed in \cite{Bartelmann2015a}, we transform the action to the new time coordinate $\eta$ defined in \eqref{eq:tau_time_definition} leading to the new Lagrange function
\begin{equation}
 \tilde{L}(\vq, \vq^{\,\prime},\eta) = \frac{m}{2} a^2 Hf \left(\tder{\vq}{\eta}\right)^2 - \frac{m}{Hf} \phi(\vq,t) \; \rightarrow \; \vp_{\mathrm{can}} = m a^2 Hf \vq^{\,\prime} \;.
\label{eq:A-02}
\end{equation}
We insert the corresponding Hamiltonian equations of motion in terms of the canonical momentum $\vp_{\mathrm{can}}$ into the generating functional and then rescale to a new momentum $\vp$ such that the generating functional is invariant. In terms of this new momentum $\vp$ we have the following equations of motion,
\begin{equation}
 \vq^{\,\prime} = \vp \coloneqq \frac{\vp_{\mathrm{can}}}{m a^2 Hf} = \frac{\dot{\vq}}{Hf} \;, \qquad \vp^{\,\prime} = - \left( \frac{3}{2} \frac{\Omega_\mm}{f^2} - 1 \right) \vp - \nabla_q \, \tilde{\phi} \;,
\label{eq:A-03}
\end{equation}
where the modified potential now obeys Poisson's equation in the form \eqref{eq:poisson_equation}. For the fiducial $\Lambda$CDM cosmology we are interested in, $\Omega_\mm / f^2 \approx 1$ is a very good approximation from the matter-dominated epoch onward \cite{Bernardeau2002,Matarrese}. Setting the potential force in \eqref{eq:A-03} to zero, we arrive at the free equation of motion
\begin{equation}
 (\partial_\eta + F) \, \vx = 0 \;, \quad F = \matrix{cc}{ 0_3 & -\mathcal{I}_3 \\ 0_3 & \frac{1}{2} \mathcal{I}_3 } \;, \quad \mathcal{G}(\eta_1,\eta_2) = \nexp{-\bsi{\eta'}{\eta_2}{\eta_1} F } \;,
\label{eq:A-04}
\end{equation}
where the retarded Green's function $\mathcal{G}$ \cite{Bartelmann2015a} is the propagator \eqref{eq:KFT_particle_propagator_cosmic_solution} for free particle motion in phase-space.

We already derived the appropriate set of SPT equations \eqref{eq:SPT_continuity_equation_with_velocity_field}--\eqref{eq:SPT_poisson_equation_massless_potential} for describing CDM as fluid. We adopt the SSA by setting $\sigma^{ij} = 0$ and introduce deviations from the mean cosmic background,
\begin{equation}
 \rho_m(\vq,t) = m \mpd(t) \bigl( 1 + \delta(\vq,t) \bigr) \;, \qquad \vec{v}(\vec{r},t) = H(t) \, \vec{r} + a(t) \, \vec{v}_{\mathrm{pec}}(\vq,t) \;.
\label{eq:A-05}
\end{equation}
Again we change the time coordinate to \eqref{eq:tau_time_definition} and rescale the peculiar velocity field to $\vupec = \vec{v}_{\mathrm{pec}}/Hf$. After subtracting the homogeneous background evolution from the equations of motion we find
\begin{align}
 \partial_\eta \delta + \nabla_q \cdot \left( \left(1+\delta\right) \vupec \right) &= 0 \;, \label{eq:A-06a} \\
 \partial_\eta \vupec + \left(\frac{3}{2} \frac{\Omega_\mm}{f^2} - 1 \right) \vupec + \left( \vupec \cdot \nabla_q \right) \vupec + \nabla_q \tilde{\phi} &= 0 \;, \label{eq:A-06b} \\
 \nabla_q^2 \tilde{\phi} &= \frac{3}{2} \frac{\Omega_\mm}{f^2} \delta \label{eq:A-06c} \;.
\end{align}
After introducing the $\varphi_a$ field doublet notation \eqref{eq:SPT_field_doublet}, we take the divergence of Euler's equation, insert Poisson's equation into the result and transform to Fourier space, to find the compact form \cite{Scoccimarro2001}
\begin{equation}
 \left(\delta_{ab} \partial_{\eta_1} + \Omega_{ab} \right) \varphi_b(1) - \usi{2} \usi{3} \gamma_{abc}\left(1,-2,-3\right) \varphi_b(2) \varphi_c(3) = 0 \;.
\label{eq:A-07}
\end{equation}

The linear equations of motion are defined by the matrix
\begin{equation}
 \Omega_{ab} = \matrix{cc}{ 0 & -1 \\ -\frac{3}{2} \frac{\Omega_\mm}{f^2} & \; \left(\frac{3}{2} \frac{\Omega_\mm}{f^2} - 1 \right) } \quad \rightarrow \quad \Omega_{ab}^{(0)} = \matrix{cc}{ 0 & \; -1 \\ 0 & \; 1/2 } \;,
\label{eq:A-08}
\end{equation}
where the second expression holds in the regime where gravity has been switched off and we used the approximation $\Omega_\mm / f^2 \approx 1$. Since $\Omega_{ab}^{(0)}$ is the two-dimensional analogue of $F$ in \eqref{eq:A-04} we can directly read off the linear propagator \eqref{eq:SPT_field_propagator_free_solution} from the KFT result. This is no surprise since in the absence of interactions the linearised version of \eqref{eq:A-06b} describes a velocity field whose evolution at any single point in space is decoupled from its surroundings due to the absence of spatial derivatives. It must thus describe inertial motion of fluid elements, and since $\vupec$ has the same dimensions as the particle momenta in KFT, it has to follow the same equation of motion. The fact that $\delta$ has the same time evolution as a particle position is then simply a consequence of the linearised continuity equation \eqref{eq:A-06a}.

The non-linear terms are contained in the vertex tensor
\begin{equation}
\gamma_{abc}(1,2,3) = (2\pi)^3 \dirac\bigl(\vk_1 + \vk_2 + \vk_3\bigr) \, \dirac(t_1 - t_2) \, \dirac(t_1 - t_3) \, \tilde{\gamma}_{abc}(1,2,3) \;,
\label{eq:A1-09}
\end{equation}
which has the well-known non-vanishing entries \cite{Bernardeau2002, Matarrese}
\begin{alignat}{2}
  \tilde{\gamma}_{121}(1,2,3) &= \frac{\alpha(\vk_2,\vk_3)}{2} &&\coloneqq \frac{\bigl(\vk_2 + \vk_3\bigr) \cdot \vk_2}{2 \vk_2^2} \;, \label{eq:A1-10a} \\
  \tilde{\gamma}_{112}(1,2,3) &= \frac{\alpha(\vk_3,\vk_2)}{2} &&\coloneqq \frac{\bigl(\vk_3 + \vk_2\bigr) \cdot \vk_3}{2 \vk_3^2} \;, \label{eq:A1-10b} \\
  \tilde{\gamma}_{222}(1,2,3) &= \beta(\vk_2,\vk_3) &&\coloneqq \frac{\bigl(\vk_2 + \vk_3\bigr)^2 \bigl(\vk_2 \cdot \vk_3\bigr)}{2 \vk_2^2 \vk_3^2} \;. \label{eq:A1-10c} 
\end{alignat}
They are related to one another by
\begin{equation}
 2 \, \alpha(\vk_1, \vk_2) \, \alpha(\vk_2, \vk_1) = \alpha(\vk_1, \vk_2) + \alpha(\vk_2, \vk_1) + 2 \, \beta(\vk_1, \vk_2) \;.
\label{eq:A1-11}
\end{equation}
This will be useful later in order to show that the one-loop mode coupling contributions to the density power spectrum are actually the same in SPT and KFT.

The statistical evolution of the fluid can be encoded in the generating functional \eqref{eq:SPT_generating_functional}. The initial probability distribution $\mathcal{P}(\varphi_a^{(\mi)})$ is chosen as a zero-mean Gaussian with Fourier space covariance matrix
\begin{equation}
\begin{split}
 P_{ab}^{(\mi)}(1,2) &= \matrix{cc}{ \ave{\varphi^{(\mi)}_\delta(\vk_1) \, \varphi^{(\mi)}_\delta(\vk_2)} & \; \ave{\varphi^{(\mi)}_\delta(\vk_1) \, \varphi^{(\mi)}_\theta(\vk_2)} \\ \ave{\varphi^{(\mi)}_\theta(\vk_1) \, \varphi^{(\mi)}_\delta(\vk_2)} & \; \ave{\varphi^{(\mi)}_\theta(\vk_1) \, \varphi^{(\mi)}_\theta(\vk_2)} } \\
 &= (2\pi)^3 \dirac\bigl(\vk_1 + \vk_2\bigr) P^{(\mi)}_\delta(k_1) \matrix{cc}{ 1 & 1 \\ 1 & 1 } \;,
\end{split}
\label{eq:A1-12}
\end{equation}
where the second line is obtained by combining \eqref{eq:SPT_field_doublet}, \eqref{eq:growing_mode_ini_condition2} and \eqref{eq:A2-01}. Now the path integrals in \eqref{eq:SPT_generating_functional} can be executed and according to \cite{Matarrese} we find
\begin{equation}
 Z_{\textsc{SPT}}[J,K] = \e^{\mi \hat{\chi}_a \gamma_{abc} \hat{\varphi}_b \hat{\varphi}_c} \; \e^{- \frac{1}{2} J_a P_{ab}^{\mathrm{lin}} J_b - \mi J_a g_{ab} K_b} \;,
\label{eq:A1-13}
\end{equation}
where repeated indices imply summation over field types and integration over arguments. The linearly evolved power spectrum matrix is obtained by simply evolving the initial $P_{ab}^{(\mi)}$ \eqref{eq:A1-12} with the linear propagator \eqref{eq:SPT_field_propagator_free_solution}, leading to
\begin{equation}
  P_{ab}^{\mathrm{lin}}(1,2) = (2\pi)^3 \dirac\bigl(\vk_1 + \vk_2\bigr) P^{(\mi)}_\delta(k_1) \, g_{ac}(\eta_1,0) \, g_{bd}(\eta_2,0) \, \matrix{cc}{ 1 & 1 \\ 1 & 1 } \;.
\label{eq:A1-14}
\end{equation}
Expanding the two exponentials and applying the appropriate functional derivatives \wrt $J_a$, to obtain correlators of $\varphi_a$, then leads to to the Feynman diagrams shown in \eqref{eq:SPT_diagram_building_blocks} \cite{Matarrese, Crocce2006, Crocce2006a}.

\section{Expansion of the KFT power spectrum} \label{AppendixB}

The power spectrum $\iniPS$ of the initial density contrast is defined in the usual way as the Fourier transform of the correlation function,
\begin{equation}
 C_{\delta_1 \delta_2}\bigl(|\vqi_{12}|\bigr) = \ave{\delta^{(\mi)}_1 \delta^{(\mi)}_2} = \int_{h} \, e^{\mi \, \vh \cdot \vqi_{12}} \, \iniPS(h),
\label{eq:A2-01}
\end{equation}
where $\delta^{(\mi)}_j = \delta^{(\mi)}(\vqi_j)$ and $\vqi_{12} = \vqi_{1} - \vqi_{2}$. We combine \eqref{eq:growing_mode_ini_condition2} with the assumption that $\vupeci$ is irrotational in order to obtain
\begin{equation}
 \vupeci(\vk) = \frac{\mi \vk}{k^2} \, \ini{\delta}(\vk) \;.
\label{eq:A2-02}
\end{equation}
As $\vupeci$ takes the role of the initial momentum field $\vini{P}$ in \eqref{eq:KFT_conditioned_sampling_distribution}, we can write density-momentum cross-correlation and momentum auto-correlation as
\begin{alignat}{2}
 \vec{C}_{p_1 \delta_2}\bigl(|\vqi_{12}|\bigr) &= \ave{\vec{u}_{\mpec,1} \, \ini{\delta}_2} &&= \int_{h} \, e^{\mi \, \vh \cdot \vqi_{12}} \, \frac{\mi \vh}{h^2} \, \iniPS(h) \label{eq:A2-03a} \;,\\
 C_{p_1 p_2}\bigl(|\vqi_{12}|\bigr) &= \ave{\vec{u}_{\mpec,1} \otimes \vec{u}_{\mpec,2}} &&= \int_{h} \, e^{\mi \, \vh \cdot \vqi_{12}} \, \frac{\vh \otimes \vh}{h^4} \, \iniPS(h) \;. \label{eq:A2-03b}
 \end{alignat}
 
We now expand \eqref{eq:KFT_exact_free_powerspectrum} up to second order in $\iniPS$. The expansion of the Fourier integral can be expressed in terms of the two functions
\begin{equation}
 A_1 \coloneqq \alpha\bigl(\vk_1 - \vh, \vh\bigr) = \frac{\vk_1 \cdot \vh}{h^2} \;, \quad A_2 \coloneqq \alpha\bigl(\vh, \vk_1 - \vh\bigr) = \frac{\vk_1 \cdot \bigl(\vk_1 - \vh\bigr)}{\bigl(\vk_1 -\vh\bigr)^2} \;.
\label{eq:A2-04}
\end{equation}
The exponential by itself gives
\begin{equation}
\begin{split}
 &\int_{\qin_{12}} \e^{-\mi \vk_1 \cdot \vqi_{12}} \left( \e^{\mg_1 \, \mg_2 \, \vk^{\,\top}_1 C_{p_1 p_2} \vk_1} - 1 \right) \\
 \xrightarrow[\leq \mathcal{O}\bigl(P_\delta^{(\mi) \, 2}\bigr)]{} & \, \mg_1 \, \mg_2 \, \iniPS(k_1)  + \frac{\mg_1^2 \, \mg_2^2}{2} \int_h \, A_1^2 \, A_2^2 \, \iniPS(h) \, \iniPS(\Delta h)  \;,
\end{split}
 \label{eq:A2-05}
\end{equation}
where $\Delta\vh \eqqcolon \vk_1 - \vh$. The combination of $C_{\delta_1 \delta_2}$ with the exponential results in
\begin{equation}
 \int_{\qin_{12}} \e^{-\mi\vk_1 \cdot \vqi_{12}} \, C_{\delta_1 \delta_2} \, \e^{\mg_1 \, \mg_2 \, \vk^{\,\top}_1 C_{p_1 p_2} \vk_1} \xrightarrow[\leq \mathcal{O}\bigl(P_\delta^{(\mi) \, 2}\bigr)]{} \iniPS(k_1) + 
   \mg_1 \mg_2 \int_h \, A_1^2 \, \iniPS(h) \, \iniPS(\Delta h) \;.
\label{eq:A2-06}
\end{equation}
Similarly, combining the $\vec{C}_{p_1 \delta_2}$ term with the exponential gives
\begin{equation}
\begin{split}
  -\mi &\left(\mg_1 + \mg_2\right) \int_{\qin_{12}} \, \e^{-\mi\vk_1 \cdot \vqi_{12}} \, \vk_1  \cdot  \vec{C}_{p_1 \delta_2} \,  \e^{\mg_1 \, \mg_2 \, \vk^{\,\top}_1 C_{p_1 p_2} \vk_1} \\
  \xrightarrow[\leq \mathcal{O}\bigl(P_\delta^{(\mi) \, 2}\bigr)]{} \, & \left(\mg_1 + \mg_2\right) \left( \iniPS(k_1) + \mg_1 \mg_2 \int_h \, A_1 \, A_2^2 \, \iniPS(h) \, \iniPS(\Delta h)  \right) \;.
\end{split}
\label{eq:A2-07}
\end{equation}
The quadratic $\vec{C}_{p_1 \delta_2}$ term is already of second order in $\iniPS$, so its combination with the exponential leaves
\begin{equation}
\begin{split}
  -\mg_1 \mg_2 &\int_{\qin_{12}} \e^{-\mi\vk_1 \cdot \vqi_{12}} \bigl(\vk_1 \cdot \vec{C}_{p_1 \delta_2}\bigr)^2 \,\e^{\mg_1 \, \mg_2 \, \vk^{\,\top}_1 C_{p_1 p_2} \vk_1} \\
  &\xrightarrow[\leq \mathcal{O}\bigl(P_\delta^{(\mi) \, 2}\bigr)]{} \mg_1 \mg_2 \int_h \, A_1 \, A_2 \, \iniPS(h) \, \iniPS(\Delta h)  \;.
\end{split}
\label{eq:A2-08}
\end{equation}

Collecting the linear terms immediately leads to \eqref{eq:KFT_linear_power_spectrum}. The quadratic self-energy contribution \eqref{eq:KFT_SE_term} is obtained by combining these linear terms with the first-order expansion of the Gaussian damping factor in \eqref{eq:KFT_exact_free_powerspectrum}. In order to rewrite the remaining quadratic mode-coupling terms in the form \eqref{eq:KFT_mode_coupling_term}, we further introduce $A_3 \coloneqq 2 \beta\bigl(\vk_1 - \vh, \vh\bigr)$. From \eqref{eq:A1-11} we see that $2 A_1 A_2 = A_1 + A_2 + A_3$. Applying the substitution $\vh \rightarrow \vk_1 - \vh$ to the above integrals, the only change in their form is $A_1 \leftrightarrow A_2, A_3 \rightarrow A_3$. Finally, we add the four loop integrals in \eqref{eq:A2-05}--\eqref{eq:A2-08} and group them into three terms according to their scaling with the free propagators. Under the integrals we can then use the above relations to find 
\begin{equation}
\begin{split}
 A_1^2 + A_1 A_2 = \frac{1}{2} \left(A_1^2 + 2A_1 A_2 + A_2^2\right) = \frac{1}{2} \left(A_1 + A_2\right)^2 &= 2 F_{\textsc{A}}^2 \;, \\
 A_1 A_2^2 = A_1 A_2 \frac{1}{2}\left(A_2 + A_2\right) = \frac{1}{2}\left(A_1 + A_2 + A_3\right) \frac{1}{2}\left(A_1 + A_2\right) &= F_{\textsc{A}} F_{\textsc{B}} \;, \\
 A_1^2 A_2^2 = \frac{1}{4} \left(A_1 + A_2 + A_3\right)^2 &= F_{\textsc{B}}^2 \;,
\end{split}
 \label{eq:A2-09}
\end{equation}
where the kernels $F_{\textsc{A}}, F_{\textsc{B}}$ defined in \eqref{eq:FA_kernel_definition}, \eqref{eq:FB_kernel_definition} are evaluated at $\vk_1 \rightarrow \Delta\vh, \vk_2 \rightarrow \vh$. These finally lead back to \eqref{eq:KFT_mode_coupling_term}.

\section{The KFT bispectrum} \label{App3}

The non-interacting density contrast bispectrum in KFT is given by connected three-particle correlations, which correspond to connected three-particle correlation diagrams in the language introduced in \cite{Fabis2018}. The general form of the bispectrum is given by
\begin{equation}
  {}^{\textsc{KFT}}G^{(0)}_{\delta \delta \delta}(1,2,3) = \mpd^3 (2\pi)^3 \dirac\bigl(\vk_1 + \vk_2 + \vk_3\bigr) \e^{-\frac{\sigma_p^2}{2} \left( \mg_1^2 \, k_1^2 + \mg_2^2 \, k_2^2 + \mg_3^2 \, k_3^2\right)} \; \sum_{\mathclap{\{1,2,3\}_{\{A,B,C\}}}} \, \mathcal{C}_{\mathrm{con}}^{(3)} \;.
\label{eq:A3-01}
\end{equation}
The sum of all connected three-particle diagrams $\mathcal{C}_{\mathrm{con}}^{(3)}$ must itself be summed over all possibilities to assign the three non-empty sets $A=\{1\}$, $B=\{2\}$, $C=\{3\}$ of \textit{field labels} to the three \textit{particle labels} $1,2,3$. Since we are only interested in terms of up to second order in $\iniPS$, we only need to consider the following seven correlation diagrams containing two lines,
\begin{equation}
 \mathcal{C}^{(3)}_{\mathrm{con}}\Big|_{2 \, \mathrm{lines}} = \, \mytikz[2.0ex]{\threepattern[1][2][3] \ddline{p1}{p3} \dpline{p2}{p3}}
 + \mytikz[2.0ex]{\threepattern[1][2][3] \ddline{p1}{p3} \ppline{p2}{p3}} 
 +\mytikz[2.0ex]{\threepattern[1][2][3] \dpline{p3}{p1} \dpline{p2}{p3}}
 + \mytikz[2.0ex]{\threepattern[1][2][3] \dpline{p1}{p3} \ppline{p2}{p3}} 
 + \mytikz[2.0ex]{\threepattern[1][2][3] \dpline{p3}{p1} \ppline{p2}{p3}}
 + \frac{1}{2} \mytikz[2.0ex]{\threepattern[1][2][3] \dpline{p1}{p3} \dpline{p2}{p3}}
 + \frac{1}{2} \mytikz[2.0ex]{\threepattern[1][2][3] \ppline{p1}{p3} \ppline{p2}{p3}} \;,
\label{eq:A3-02}
\end{equation}
where the numbers label the three particles. The linear contributions of each individual line are given by
\begin{align}
 \mytikz{\hortwopattern[1_A][3_C] \ddline{p1}{p2} } &= \int_{q^{(\mi)}_{13}} \e^{-\mi\vk_A \cdot \vqi_{13}} \, C_{\delta_1 \delta_3} \, \e^{-\mg_A \, \mg_C \, \vk^{\,\top}_A C_{p_1 p_3} \vk_C} \xrightarrow[\leq \mathcal{O}({\iniPS})]{} \iniPS(k_A)  \label{eq:A3-03a} \;, \\
 \mytikz{\hortwopattern[1_A][3_C] \dpline{p2}{p1} } &= \int_{q^{(\mi)}_{13}} \e^{-\mi\vk_A \cdot \vqi_{13}} \, \bigl( -\mi \mg_A \vk_A \cdot \vec{C}_{p_1 \delta_3} \bigr) \, \e^{-\mg_A \, \mg_C \, \vk^{\,\top}_A C_{p_1 p_3} \vk_C} \xrightarrow[\leq \mathcal{O}({\iniPS})]{} \mg_A \iniPS(k_A)  \;, \label{eq:A3-03b} \\
 \begin{split}
 \mytikz{\hortwopattern[1_A][3_C] \dpline{p1}{p2} } &= \int_{q^{(\mi)}_{13}} \e^{-\mi\vk_A \cdot \vqi_{13}} \, \bigl( -\mi \mg_C \vk_C \cdot \vec{C}_{\delta_1 p_3} \bigr) \, \e^{-\mg_A \, \mg_C \, \vk^{\,\top}_A C_{p_1 p_3} \vk_C} \\
 &\quad \xrightarrow[\leq \mathcal{O}({\iniPS})]{} -\mg_C \frac{\vk_C \cdot \vk_A}{k_A^2} \iniPS(k_A)  \;,
 \end{split}
 \label{eq:A3-03c} \\
 \mytikz{\hortwopattern[1_A][3_C] \ppline{p1}{p2} } &= \int_{q^{(\mi)}_{13}} \e^{-\mi\vk_A \cdot \vqi_{13}} \left( \e^{-\mg_A \, \mg_C \, \vk^{\,\top}_A C_{p_1 p_3} \vk_C} - 1\right) \xrightarrow[\leq \mathcal{O}({\iniPS})]{} -\mg_A \mg_C \frac{\vk_C \cdot \vk_A}{k_A^2} \iniPS(k_A)  \;.\label{eq:A3-03d}
\end{align}

For each diagram the sum over assignments of sets $A,B,C$ to the particles $1,2,3$ leads to $3! = 6$ contributions. The strategy is to combine the contributions from the assignments $\{1_A,2_B,3_C\}$ and $\{1_B, 2_A, 3_C\}$ into seven sums of two diagrams each. The full contribution can then be written as the sum over cyclic permutations of these. The first four terms are
\begin{align}
 \mytikz[2.0ex]{\threepattern[1_A][2_B][3_C] \ddline{p1}{p3} \dpline{p2}{p3}} + \mytikz[2.0ex]{\threepattern[1_B][2_A][3_C] \ddline{p1}{p3} \dpline{p2}{p3}} &= -\iniPS(k_A) \, \iniPS(k_B) \, \mg_C \, \vk_C \left(\frac{\vk_A}{k_A^2} + \frac{\vk_B}{k_B^2} \right) \;, \label{eq:A3-04a} \\
 \mytikz[2.0ex]{\threepattern[1_A][2_B][3_C] \ddline{p1}{p3} \ppline{p2}{p3}} + \mytikz[2.0ex]{\threepattern[1_B][2_A][3_C] \dpline{p3}{p1} \dpline{p2}{p3}} &= -\iniPS(k_A) \, \iniPS(k_B) \, \mg_B \, \mg_C \, \vk_C \left(\frac{\vk_A}{k_A^2} + \frac{\vk_B}{k_B^2} \right) \;, \label{eq:A3-04b}  \\
 \mytikz[2.0ex]{\threepattern[1_A][2_B][3_C] \dpline{p3}{p1} \dpline{p2}{p3}} + \mytikz[2.0ex]{\threepattern[1_B][2_A][3_C] \ddline{p1}{p3} \ppline{p2}{p3}} &= -\iniPS(k_A) \, \iniPS(k_B) \, \mg_A \, \mg_C \, \vk_C \left(\frac{\vk_A}{k_A^2} + \frac{\vk_B}{k_B^2} \right) \;, \label{eq:A3-04c}  \\
 \mytikz[2.0ex]{\threepattern[1_A][2_B][3_C] \dpline{p3}{p1} \ppline{p2}{p3}} + \mytikz[2.0ex]{\threepattern[1_B][2_A][3_C] \dpline{p3}{p1} \ppline{p2}{p3}} &= -\iniPS(k_A) \, \iniPS(k_B) \, \mg_A \, \mg_B \, \mg_C \, \vk_C \left(\frac{\vk_A}{k_A^2} + \frac{\vk_B}{k_B^2} \right) \;. \label{eq:A3-04d} 
\end{align}
We sum all four terms, use the leading Dirac delta distribution in \eqref{eq:A3-01} to replace $\vk_C = -\bigl(\vk_A + \vk_B\bigr)$, and with \eqref{eq:FA_kernel_definition} find that 
\begin{equation}
\begin{split}
 &\iniPS(k_A) \, \iniPS(k_B) \, \mg_C \, (1 + \mg_A + \mg_B + \mg_A \, \mg_B) \, \bigl(\vk_A + \vk_B\bigr) \left(\frac{\vk_A}{k_A^2} + \frac{\vk_B}{k_B^2} \right) \\
 &\qquad= \iniPS(k_A) \, \iniPS(k_B) \, \mg_C \, (1+\mg_A) \, (1+\mg_B) \, 2 F_{\textsc{A}}\bigl(\vk_A,\vk_B\bigr) \;.
\end{split}
 \label{eq:A3-05}
\end{equation}
The remaining three diagrams are combined into
\begin{align}
 \frac{1}{2} \, \mytikz[2.0ex]{\threepattern[1_A][2_B][3_C] \dpline{p1}{p3} \dpline{p2}{p3}} + \frac{1}{2} \, \mytikz[2.0ex]{\threepattern[1_B][2_A][3_C] \dpline{p1}{p3} \dpline{p2}{p3}} &= \iniPS(k_A) \, \iniPS(k_B) \, \mg_C^2 \, \frac{\vk_C \cdot \vk_A}{k_A^2} \, \frac{\vk_C \cdot \vk_B}{k_B^2} \;, \label{eq:A3-06a} \\ 
 \mytikz[2.0ex]{\threepattern[1_A][2_B][3_C] \dpline{p1}{p3} \ppline{p2}{p3}} + \mytikz[2.0ex]{\threepattern[1_B][2_A][3_C] \dpline{p1}{p3} \ppline{p2}{p3}} &= \iniPS(k_A) \, \iniPS(k_B) \left(\mg_A + \mg_B\right) \mg_C^2 \, \frac{\vk_C \cdot \vk_A}{k_A^2} \, \frac{\vk_C \cdot \vk_B}{k_B^2} \;, \label{eq:A3-06b} \\
 \frac{1}{2} \, \mytikz[2.0ex]{\threepattern[1_A][2_B][3_C] \ppline{p1}{p3} \ppline{p2}{p3}} + \frac{1}{2} \, \mytikz[2.0ex]{\threepattern[1_B][2_A][3_C] \ppline{p1}{p3} \ppline{p2}{p3}} &= \iniPS(k_A) \, \iniPS(k_B) \, \mg_A \, \mg_B \, \mg_C^2 \, \frac{\vk_C \cdot \vk_A}{k_A^2} \, \frac{\vk_C \cdot \vk_B}{k_B^2} \;. \label{eq:A3-06c} 
\end{align}
We sum these three terms, again use $\vk_C = -\bigl(\vk_A + \vk_B\bigr)$, insert \eqref{eq:A1-11} and \eqref{eq:FB_kernel_definition}, and find
\begin{equation}
\begin{split}
  &\iniPS(k_A) \, \iniPS(k_B) \, \mg_C^2 \bigl(1 + \mg_A + \mg_B + \mg_A \, \mg_B\bigr) \frac{ \bigl(\vk_A + \vk_B\bigr) \cdot \vk_A}{k_A^2} \, \frac{\left(\vk_A + \vk_B\right)\cdot \vk_B}{k_B^2} \\
 &\qquad= \iniPS(k_A) \, \iniPS(k_B) \, \mg_C^2 \left(1+\mg_A\right) \bigl(1+\mg_B\bigr) F_{\textsc{B}}\left(\vk_A,\vk_B\right) \;.
\end{split}
 \label{eq:A3-07}
\end{equation}
Adding \eqref{eq:A3-05} and \eqref{eq:A3-07}, summing all cyclic permutations of the label assignments and inserting the result into \eqref{eq:A3-01} yields \eqref{eq:KFT_free_quadratic_bispectrum} if we remember that 
$A=\{1\}, B=\{2\}, C=\{3\}$.

\bibliographystyle{plain}
\bibliography{Bibliography}

\end{document}